\documentclass[%
reprint, 11pt
 amsmath,
 amssymb,
 aps, 
]{revtex4-1}

\def\bra{\langle}
\def\ket{\rangle}

\def\eps{\epsilon}
\newcommand{\imp}{d}
\newcommand{\kk}{m}
\def\sumN{\sum_{\kk=1}^N}

\usepackage{graphicx}% Include figure files
\usepackage{dcolumn}% Align table columns on decimal point
\usepackage{bm}% bold math
\usepackage{caption}
\usepackage{subcaption}
\usepackage[colorlinks]{hyperref}% add hypertext capabilities
\begin{document}

%\preprint{APS/123-QED}

\title{Impurity-directed Transport within a Finite Disordered Lattice}% Force line breaks with \\

\author{Bradley J. Magnetta}
 \email{bmagnetta@ucla.edu}
\affiliation{
 Materials Science and Engineering - UCLA,
410 Westwood Plaza
Los Angeles, California
90095-1595, USA}
\altaffiliation{{\it Current:} bradley.magnetta@yale.edu}

\author{Gonzalo Ordonez}%
 \email{gordonez@butler.edu}
\affiliation{%
Department of Physics and Astronomy,
Butler University, 4600 Sunset Ave. Indianapolis, Indiana 46208, USA
}
\author{Savannah Garmon}%
 \email{sgarmon@p.s.osakafu-u.ac.jp}
\affiliation{%
Department of Physical Science,
Osaka Prefecture University, 1-1 Gakuen-cho, Nakaku, Sakai, Osaka 599-8531, Japan
}%

\date{\today}% It is always \today, today,

\begin{abstract}
We consider a finite, disordered 1D quantum lattice with a side-attached impurity. We study theoretically the transport of a single electron from the impurity into the lattice, at zero temperature. The transport is dominated by Anderson localization and, in general, the electron motion has a random character due to the lattice disorder. However, we show that by adjusting the impurity energy the electron can attain quasi-periodic motions, oscillating between the impurity and a small region of the lattice. This region corresponds to the center of a localized state in the lattice with an energy matched by that of the impurity.  By precisely tuning the impurity energy, the electron can be set to oscillate between the impurity and a  region far from the impurity, even distances larger than the Anderson localization length.  The electron oscillations result from the interference of hybrized states, which have some resemblance to Pendry's necklace states [J. B. Pendry, J. Phys. C: Solid State Phys. \textbf{20}, 733-742 (1987)].  The dependence of the electron motion on the impurity energy gives a potential mechanism for selectively routing an electron towards different regions of a 1D disordered lattice. 

\end{abstract}

\maketitle

%\tableofcontents

\section{Introduction}

Many researchers \cite{A1,B1,A2,B2,A3,A4,A5,Sasada11, Orellana03}  have studied open (infinite) models of one-dimensional regular lattices, in which an impurity is introduced that allows for control over transport and closely related properties in the lattice.  This has led us to consider the possibility that an impurity might be used to control transport even in disordered finite  systems, with one question in mind: What type of transport would occur if the ordered lattice were replaced by a disordered one?

It is well-known that disorder in quantum systems produces Anderson localization \citep{Anderson58}. There have been numerous theoretical and experimental studies on Anderson localization \cite{Kramer,Lagendijk, Segev, Hsieh,Chomette}. For example, on the theoretical side, it has been shown that in a one-dimensional lattice with random energies at each site, all the eigenstates of the Hamiltonian are localized \citep{Landauer57,Landauer70,Erdos82}. Although this result indicates that there can be no electron conductance through an {\it infinite} one-dimensional disordered lattice, sharp resonances at the band center have been noted \citep{Lambert84,Kappus81}. Such resonances are required for electron transport. In fact, Pendry \citep{Pendry87} has shown that it is possible to transmit an electron from one end of a disordered {\it finite} lattice to the other  due to the presence of ``necklace states'' that serve as stepping-stones for the electron.
Necklace states also exist in optical systems \cite{Bertolotti,Chen}.  These states form a sub-band that can induce resonant transport similar to the energy band of an ordered lattice. Resonances of finite disordered systems coupled to infinite reservoirs have been theoretically studied in \cite{Kunz,Feinberg}.

Extrapolating from these previous studies, here we consider finite disordered lattices (or quantum wires) with a side-attached impurity (``T-junction''). The impurity can be realized using a quantum dot, which constitutes a nano-control device. The properties of the dot can be altered through a gate potential allowing an experimentalist control over electron transport.  Varying the gate potential on the dot can be used to 
%``scan through'' and investigate the spectral properties of the lattice. 
probe the spectrum and localization properties of the lattice.
As we will show, we can indeed use an impurity to direct transport within a disordered lattice. In our theoretical study we will consider the case of zero temperature. Therefore the transport we will discuss is different from variable-range hopping \cite{PALee, Bosisio2}, which occurs at non-zero temperature. We will discuss possible extension of our work to the case of nonzero temperature in section \ref{sec:disc}. 
Note that our lattice is finite, but large enough so that  boundary effects only play a minor role.

Experimentally, effective 1-D systems can be synthesized by a variety of techniques \cite{A6, A7}, including lattice geometries that incorporate a side-attached quantum dot \cite{A8, A9}.  Randomized site potentials in a finite lattice might be obtained, for example, by varying segment lengths (i.e., growth times) in GaAs/GaP superlattices assembled by laser-assisted catalytic growth \cite{Chomette,A7}.  An effective side-attached dot could then potentially be introduced by doping one such segment.

%Thus, the T-junction model we will consider consists of a one-dimensional disordered lattice with a side-attached impurity.  The lattice is finite, but large enough so that  boundary effects are negligible.  The non-disordered version of the T-junction model \cite{A1} has previously been used to study resonance phenomena and transport properties \cite{Sasada11, Orellana03,A2,A3}, as well as the optical absorption spectrum \cite{A4}, and resonance influence on entanglement entropy scaling \cite{A5}.  

%In the case of a disordered lattice, we include the side-attached impurity with a variable potential as a means to probe the spectrum and localization properties of the lattice.  

While as far as we are aware there have been no studies on disordered models of electron transport incorporating a side-attached impurity, at least one experimental realization of a system similar to ours has been reported in Ref. \cite{Bliokh} using microwaves instead of electrons. The system in Ref. \cite{Bliokh} consists of a waveguide with random blocks (analogous to our disordered lattice) and an air-gap in the middle (analogous to our impurity site). The focus of Ref. \cite{Bliokh} however  is different from the focus of our present study, as we will discuss below.  In another relevant work, boundary effects on localization properties have recently been studied in finite, weakly disordered optical waveguide arrays \cite{Thompson14}. Moreover, Refs. \cite{Bosisio2,Bosisio1} have considered control of thermopower using a gate potential that shifts all the lattice energies at once. 

We will consider the  motion of the electron from the impurity to the lattice and back. The initial state is that in which the electron is completely localized in the side impurity. Hereafter this state will be referred to as the unperturbed {\it impurity state}; this state is an eigenstate of the unperturbed Hamiltonian, corresponding to the case where the impurity is decoupled from the lattice. The coupling will allow the electron to transfer between the impurity and the lattice.   

We will treat the energy of the impurity as a tunable parameter. We will study how this parameter influences the transport of the electron from the impurity to the lattice, or vice versa. 

Our main finding is that for certain impurity energies the electron can jump to small regions in the lattice; these regions are localization centers of Anderson-localized states whose energy is matched by the impurity energy. These, together with the impurity state, form hybridized states that are similar to the necklace states studied by Pendry. Interference between the hybridized states induces  Rabi-like oscillations of the electron survival probability at the impurity. Hence, the electron alternates positions between the impurity and the localization centers of the lattice states hybridized with the impurity state. 

The Rabi-like oscillations occur in the vicinity of avoided crossings in the energy spectrum of the system; these avoided crossings are induced by the interaction between the impurity and the lattice. The center of the avoided crossings signals the appearance of maximally hybridized states.  Experimental observation of avoided crossings and hybridized states similar to ours was   the main focus of Ref. \cite{Bliokh} mentioned above. The new aspect of our study relative to Ref. \cite{Bliokh} is the description of the time evolution of the electron associated with these avoided crossings and the possibility of tuning the impurity energy to predictably route electrons to different regions of the lattice. In addition, we will also point out that the range of electron transport can be larger than the localization length of the hybridized states, as long as the impurity energy is precisely tuned to match the center of the avoided crossing.

We will focus our attention  on Rabi-like oscillations involving only two or three hybridized states. Oscillations involving  many hybridized states produce an erratic pattern of motion, which is less suitable for controlled transport.

The paper is organized as follows: in sections II-VI, we introduce the model and analyze the electron transport between the impurity and the lattice for a specific realization of disorder. In section VII we consider ensemble averaging and in section VIII we discuss our results.

%%%%%%%%%%%%%%%
\section{T-Junction Lattice}
\label{sec:TJ}
We consider a T-junction lattice, consisting of a disordered lattice (a finite one-dimensional chain of quantum-wells with random energy levels) and a side impurity attached to one of the wells. The impurity is introduced as a nano-control device that will enable directed electron transport between the impurity  and a lattice segment.
\begin{figure}[h!]
\begin{center}
\includegraphics[width=3.25in, angle=0]{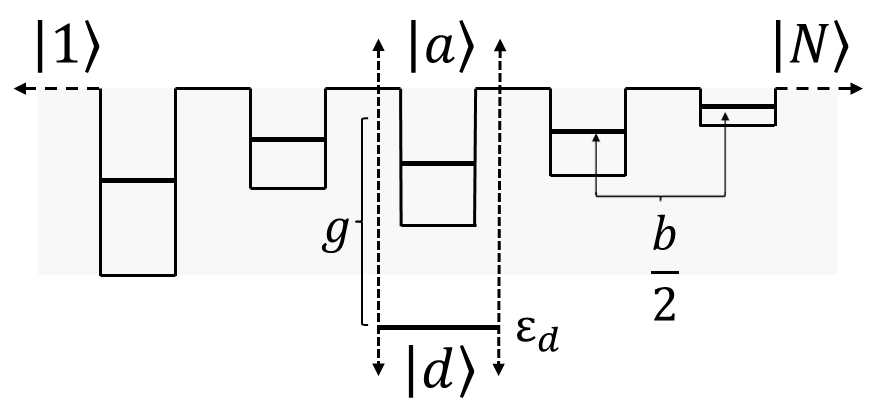} 
\end{center}
\caption{T-Junction lattice with $N$ lattice sites and impurity site $\imp$. The lattice sites have disordered energies within the range $W$ and a constant nearest neighbor interaction energy of $b/2$. The impurity has energy $\epsilon_\imp$ and is attached to the lattice at site $a$ through tunneling strength $g$. }
\label{TJL}
\end{figure}
We will focus on the motion of a single electron and will neglect Coulomb interactions altogether. We will model the lattice using a tight-binding Hamiltonian with uniform nearest-neighbor interactions, represented as a sum of lattice and impurity Hamiltonians, $H=H_{\rm lattice }+ H_{\rm \imp}$. The lattice Hamiltonian is written as
\begin{eqnarray} \label{Ham}
H_{\rm lattice}&=& \sum_{x=1}^{N} \epsilon_x |x\ket\bra x| \nonumber\\ 
&-& \frac{b}{2} \sum_{x=1}^{N-1} \left(|x+1\ket\bra x|+ |x\ket\bra x+1|\right) 
\end{eqnarray}
The energies $\epsilon_x$ are random energies uniformly distributed to introduce purely diagonal disorder. They describe unoccupied levels of the quantum wells that will roughly form an energy band. The width of the disorder $W$ is represented by the range $W=\eps_{\rm  max}-\eps_{\rm  min}$, where for simplicity we will set $\eps_{\rm  max}=W$ and $\eps_{\rm  min}=0$. Other parameters include the number of lattice sites $N$ and nearest neighbor interaction strength  $b/2$. Hereafter we will use $b$ as our energy unit. We also choose  $W=b$ such that the disorder width is comparable to the nearest-neighbor interaction strength.

The impurity Hamiltonian is given by
\begin{eqnarray}\label{eq:Himp}
H_{\rm \imp}=  \epsilon_\imp |\imp\ket\bra \imp| - g \left(|a\ket\bra \imp|  +|\imp\ket\bra a| \right)
\end{eqnarray}
The impurity is denoted as $\imp$ while the lattice attachment site is defined as site $a$, where $a\in \{1,N\}$; $\epsilon_\imp$ represents the energy of the impurity, which we treat as a tunable parameter. The impurity could be physically realized by using a quantum dot with a variable gate potential \cite{A8,A9} or by segment doping,  although the impurity energy would be fixed for an individual lattice in the latter case. Tunneling strength between the impurity and the attachment site is given by $g$.

\subsection{Characteristics of uncoupled disordered lattice}
\begin{figure}[h]
\centering
\includegraphics[width=2.5in, height=3.25in, angle=-90]{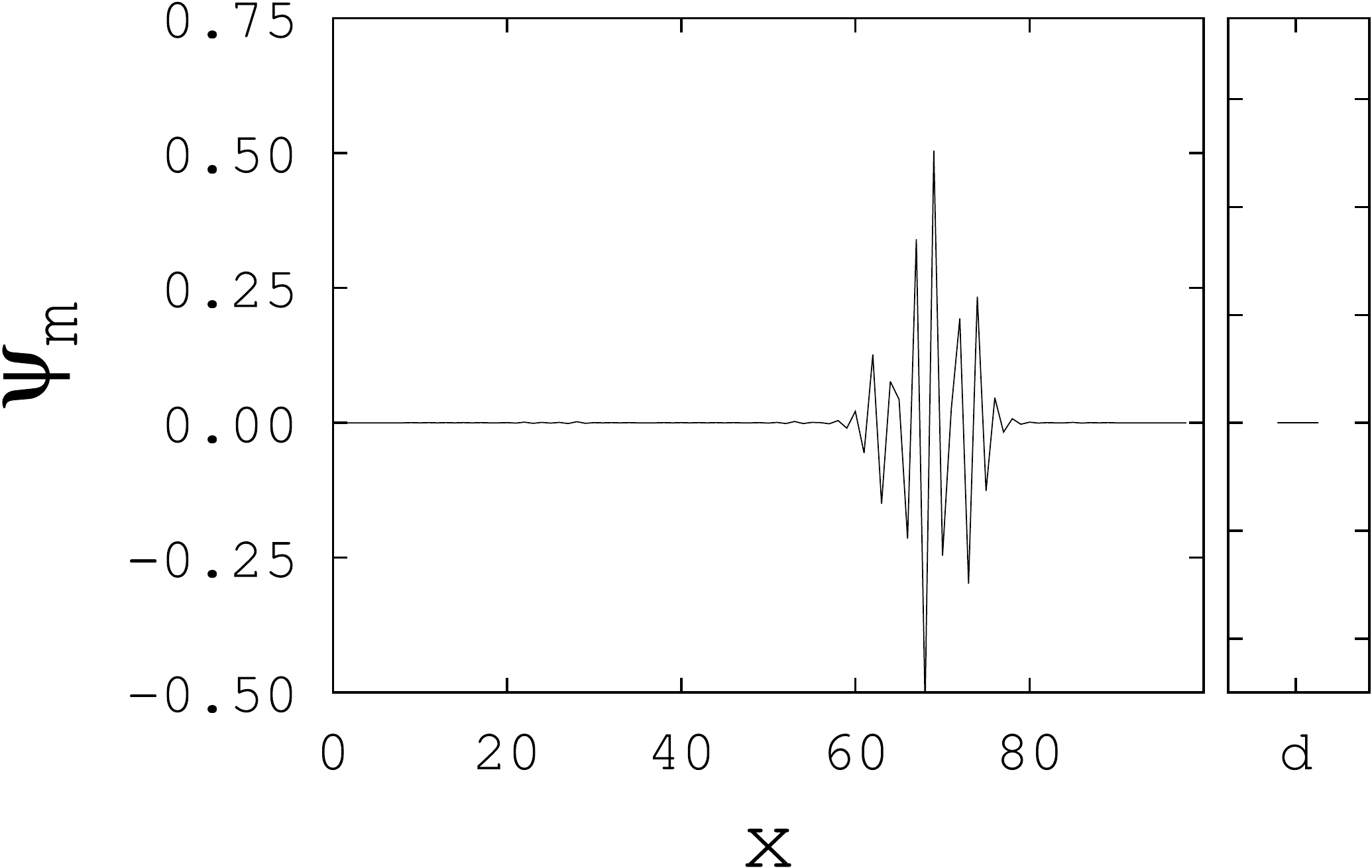}
\caption{AL lattice state with $E_\kk=-0.64$ for the uncoupled lattice. The $x$ axis is the location along the lattice, and the vertical axis is the amplitude of the state. The side bar gives the amplitude at the impurity, which in this case is zero. This state exhibits strong Anderson Localization and hence a large inverse participation number.}
\label{AL0}
\end{figure}

To better understand the capability of the side impurity to direct transport within the lattice we first investigate the influence on the spectrum of the Hamiltonian as we vary the tunneling strength. We will begin by investigating the $g=0$ case, when the lattice and impurity are uncoupled. For this case the Hamiltonian of the disconnected lattice can be diagonalized as 
\begin{eqnarray}
H_{\rm lattice}= \sum_{\kk=1}^N E_\kk |\psi_\kk\ket\bra \psi_\kk|.
\end{eqnarray}
The presence of disorder results in  Anderson Localization (AL) in the lattice. To demonstrate the occurrence of state localization we numerically diagonalized a specific realization of the lattice Hamiltonian with random site-energies. Figure \ref{AL0} shows one of the resulting localized states. In this section and in  sections~\ref{sec:avoid}-\ref{sec:Tuning} we will use this specific realization of the site energies to illustrate our results.
~ %add desired spacing between images, e. g. ~, \quad, \qquad, \hfill etc.

\begin{figure}[h]
\centering
\includegraphics[width=3.25in, height=3in]{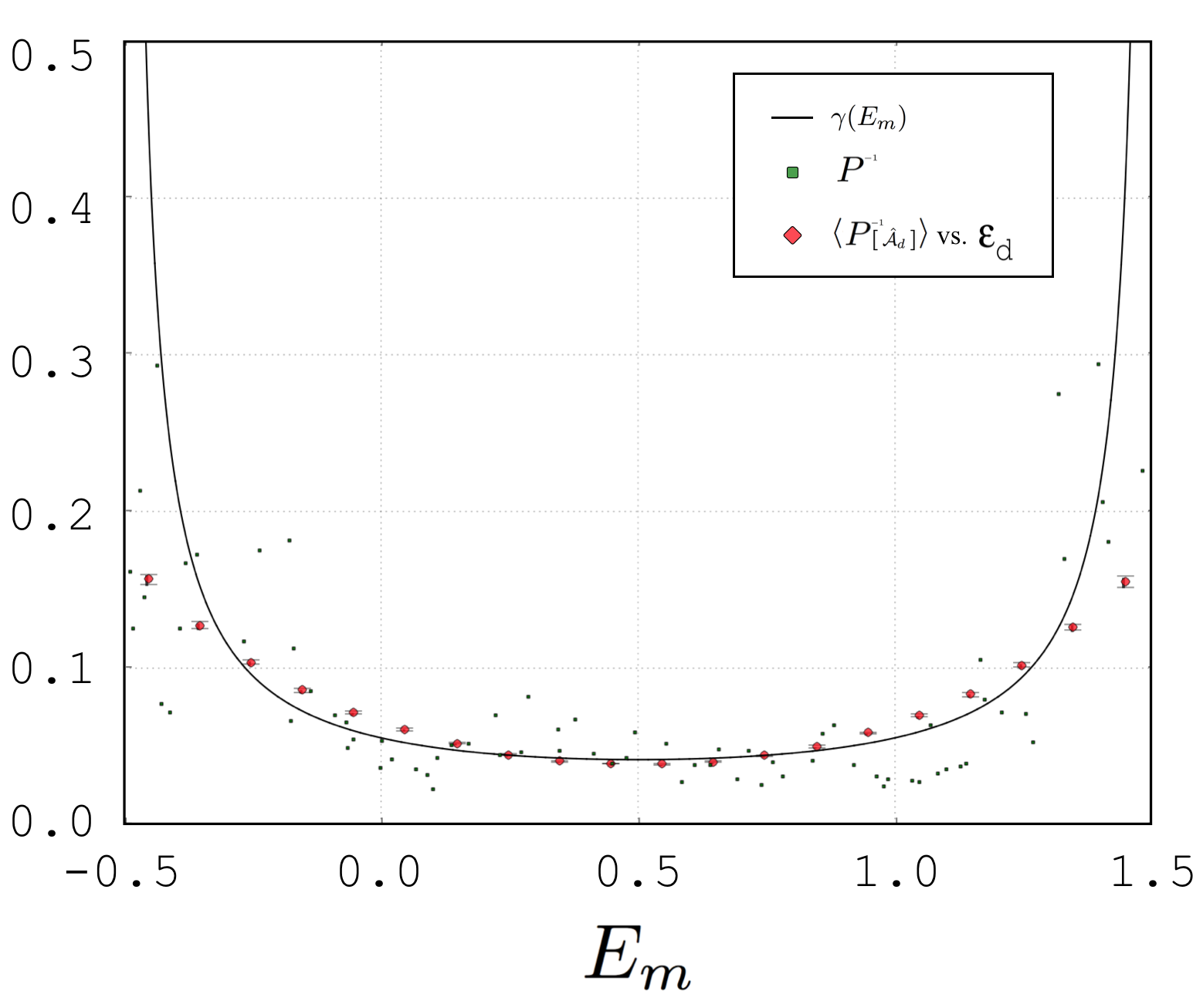}
\caption{The scattered points are the numerical inverse participation number vs. eigenvalue $E_m$ for  the uncoupled lattice eigenstates. Stronger eigenstate localization (as exhibited by the AL lattice state in figure \ref{AL0}) occurs for the eigenvalues that are closer to the edges of the lattice's energy spectrum. The solid line is the theoretical inverse localization length $\gamma(E_\kk)$ given in Eq. (\ref{gammaE}). Here we used the parameters
$b=1$ and $W=b$. The red diamonds are discussed in section VII. They represent the numerically calculated average inverse localization of hybridized states with $g=b/4$. Numerical ensemble-averaging was used with the threshold $CI=0.01\mu$. The red diamonds are plotted vs. the impurity energy $\epsilon_d$ instead of $E_m$. They approximately agree with the other two curves because AL states hybridized with the impurity satisfy $E_m \approx \epsilon_d$.  }
\label{IP}
\end{figure}%
~ %add desired spacing between images, e. g. ~, \quad, \qquad, \hfill etc.

The degree of state localization can be determined by  the second moment of probability density, the inverse participation number \citep{Wegner80}
%%%%%
\begin{eqnarray}
P^{-1}=\sum_{x}|\Psi(x)|^4,
\end{eqnarray}
%%%%
where $\Psi(x)\equiv \bra x|\Psi\ket$.
Our numerical example produced states with a variety of localization characteristics. Figure \ref{IP} illustrates this variety by plotting  state eigenvalues against their corresponding inverse participation number. 
The inverse participation number is fairly well approximated by the theoretical inverse localization length
%%%%%
\begin{eqnarray}\label{gammaE}
\gamma(E_\kk) = \frac{W^2}{24[b^2-(E_\kk-E_0)^2]}
\end{eqnarray}
%%%%
derived in Ref. \cite{Kramer}, where $E_0$ is the center of the random energies ($E_0=0.5$ here).
%Note that for $g=0$ the state $|\imp\ket$ in Eq. (\ref{eq:Himp}) represents a localized state at the impurity. The  localization of $|\mp\ket$ is not due to disorder, but simply to the fact that for $g=0$ the impurity is disconnected from the lattice. As we will show in the next section, when $g\ne 0$ the eigenstates of the Hamiltonian are expressed as a superposition involving the  impurity state and the Anderson-localized states of the lattice, forming a hybrid state.

\subsection{Coupled case} 
Having described the eigenstates of the Hamiltonian for the $g=0$ uncoupled case we analyze next the $g>0$ case as a perturbation of the uncoupled system. This allows us to understand how the impurity modifies the spectral properties of the finite disordered lattice. In terms of  the eigenstates of the Hamiltonian for the uncoupled lattice, the partially diagonalized Hamiltonian then takes the form
\begin{eqnarray}
H &=& \sum_{\kk=1}^N E_\kk |\psi_\kk\ket\bra \psi_\kk| +  \epsilon_\imp |\imp\ket\bra \imp| \nonumber\\
&-& g \sum_{\kk=1}^N \left(V_\kk^* |\psi_\kk\ket  \bra \imp|  + V_\kk |\imp\ket  \bra \psi_\kk| \right)
\end{eqnarray}
%\begin{eqnarray}\label{Ham}
%H = \sum_{\kk=1}^N E_\kk |\psi_\kk\ket\bra \psi_\kk| +  \epsilon_d |d\ket\bra d| + g \sum_{\kk=1}^N \left(V_\kk^* |\psi_\kk\ket  \bra d|  + V_\kk |d\ket  \bra \psi_\kk| \right)
%\end{eqnarray}
%%
where $V_\kk = \bra a|\psi_\kk\ket$ is the amplitude of the $\kk^{\rm th}$ eigenstate at site $a$, which determines the strength of the interaction between each mode and the impurity.  We remark that due to the completeness of the eigenstates $|\psi_\kk\ket$ we have
\begin{eqnarray}
\sum_{\kk=1}^N |V_\kk|^2 =1.
\end{eqnarray}
Therefore, if a group of modes interacts strongly with the impurity, the other modes will  interact weakly.

The eigenvalue equation for an eigenstate $|\phi_j\ket$ with real eigenvalue $z_j$ is written as
\begin{eqnarray}
H|\phi_j\ket = z_j |\phi_j\ket.
\end{eqnarray}
Writing the explicit matrix elements of the Hamiltonian  gives the set of equations
\begin{eqnarray}\label{phid}
\eps_\imp \bra \imp| \phi_j\ket - g \sumN V_\kk \bra \psi_\kk | \phi_j\ket = z_j\bra \imp |\phi_j\ket,
\end{eqnarray}
\begin{eqnarray}
-g V_{\kk'}^* \bra \imp | \phi_j\ket + E_{\kk'} \bra \psi_{\kk'} | \phi_j\ket = z_j\bra \psi_{\kk'} |\phi_j\ket.
\end{eqnarray}
Letting $\kk'=\kk$ and assuming $z_j\ne E_\kk$ for all $\kk$, we solve for $\bra\psi_{\kk} |\phi_j\ket$ as
\begin{eqnarray}
\bra \psi_{\kk} |\phi_j\ket = -\frac{1}{z_j-E_\kk}  g V_{\kk}^* \bra \imp | \phi_j\ket.
\end{eqnarray}
Substituting this into Eq. (\ref{phid}) we obtain
\begin{eqnarray}\label{zev}
z_j = \eps_\imp  + g^2 \sumN  \frac{|V_\kk|^2}{z_j-E_\kk}.
\end{eqnarray}
This equation can be written as a polynomial equation of degree $N+1$  for the $N+1$ eigenvalues of the coupled, full Hamiltonian.  The corresponding eigenstates are given by 
\begin{eqnarray}\label{phidd}
| \phi_j\ket  = |\imp\ket\bra \imp |\phi_j\ket -  g \sumN  |\psi_\kk\ket \frac{V_{\kk}^*}{z_j-E_\kk}  \bra \imp | \phi_j\ket 
\end{eqnarray}
where $\bra \imp |\phi_j\ket$ is found from the normalization condition $\bra \phi_j|\phi_j\ket = 1$, which gives
\begin{eqnarray} \label{phid2'}
 |\bra d |\phi_j\ket|^2 = \left( 1 + g^2 \sumN  \frac{|V_\kk|^2}{(z_j-E_\kk)^2} \right)^{-1}.
\end{eqnarray}
This expresses the probability to find the electron at the impurity when it is in the state $|\phi_j\ket$. By taking the derivative of Eq. (\ref{zev}) with respect to $\eps_d$ we can also write Eq. (\ref{phid2'}) as
\begin{eqnarray} \label{phid3}
 |\bra d |\phi_j\ket|^2 = \frac{\partial z_j}{\partial\eps_d}.
\end{eqnarray}
It is worth pointing out that by setting $g=0$ in Eq. (\ref{zev}), at first sight we just obtain the lone uncoupled eigenvalue $z_j=\epsilon_\imp$ for the uncoupled impurity state. But the other eigenvalues (associated with the chain) have  non-trivial $g\to 0$ limits, which are not obvious from Eq. (\ref{zev}). We can see these limits more naturally with the following rearrangement, where we pull out  a specific term corresponding to the $l^{\rm th}$ uncoupled eigenvalue
\begin{eqnarray}\label{zkev}
z_l = E_l + \frac{g^2 \left|V_l\right|^2}{ \eta_l(z_l)},
\end{eqnarray}
where
\begin{eqnarray}
\eta_l(z_l)\equiv z_l -\eps_\imp -\sum_{\kk\ne l} \frac{g^2  \left|V_\kk\right|^2}{z_l-E_\kk}.
\end{eqnarray}
Note that this equation is a polynomial equation of degree $N+1$, similar to Eq. (\ref{zev}). The difference is that Eq. (\ref{zkev}) formally reduces to the eigenvalue $E_l$ of the uncoupled lattice  when $g\to0$, whereas Eq. (\ref{zev}) reduces to the impurity eigenvalue in the same limit. The eigenstates corresponding to the eigenvalue in Eq. (\ref{zkev}) are given by 
\begin{eqnarray}\label{phikk}
|\phi_l\ket = N_l\left[ |\psi_l\ket - \frac{g V_l}{\eta_l(z_l)} \left(|\imp\ket - \sum_{\kk\ne l} |\psi_\kk\ket \frac{g V_\kk^*}{z_l-E_\kk}\right)\right],\nonumber\\
\end{eqnarray}
where
\begin{eqnarray}
N_l \equiv \left[ 1 +\frac{g^2\left|V_l\right|^2}{\eta_l^2(z_l)}\left(1+ \sum_{\kk\ne l}  \frac{g^2 \left|V_\kk\right|^2}{(z_l-E_\kk)^2}\right)\right]^{-1/2}
\end{eqnarray}
is a normalization constant. 

\subsection{Perturbation due to the coupling $g$}
 Equations (\ref{zev}) and (\ref{zkev}) demonstrate perturbation characteristics that are induced by a non-zero tunneling strength $g$ between the lattice and the impurity; they also show the dependence of the eigenvalues  on the energy value of the impurity, $\epsilon_\imp$, which we will consider as a tunable parameter in the next section.

The effect of the impurity-lattice coupling is only significant in cases where
\begin{eqnarray}\label{sigVk}
\frac{|g V_\kk|}{(z_j- E_\kk)} \sim 1,
\end{eqnarray}
in Eqs. (\ref{phidd})  or (\ref{phikk}), for at least one value of $\kk$. If this condition is not met, the impurity state remains approximately isolated from the lattice (weakly hybridized). Likewise, lattice states do not become significantly altered by the impurity's presence, so they are close to the unperturbed Anderson-localized states.

When Eq. (\ref{sigVk}) is satisfied for at least one value of $\kk$, on the other hand, the impurity state in Eq. (\ref{phidd}) will become  {\it strongly hybridized with the Anderson-localized (AL) state(s) $|\psi_\kk\ket$,} retaining some of its isolated characteristics while taking on characteristics of those AL states; conversely, the AL states will become hybridized with the impurity state. Due to the lattice-impurity coupling, AL states with large $V_\kk$ can even take on each others localization characteristics; they do so using the impurity as an intermediary as indicated by the additional $g V_\kk^*$ term in Eq. (\ref{phikk}). 

\begin{figure}[h!]
\centering
        \begin{subfigure}[h]{0.5\textwidth}
                \includegraphics[width=2in, height=3.25in, angle=-90]{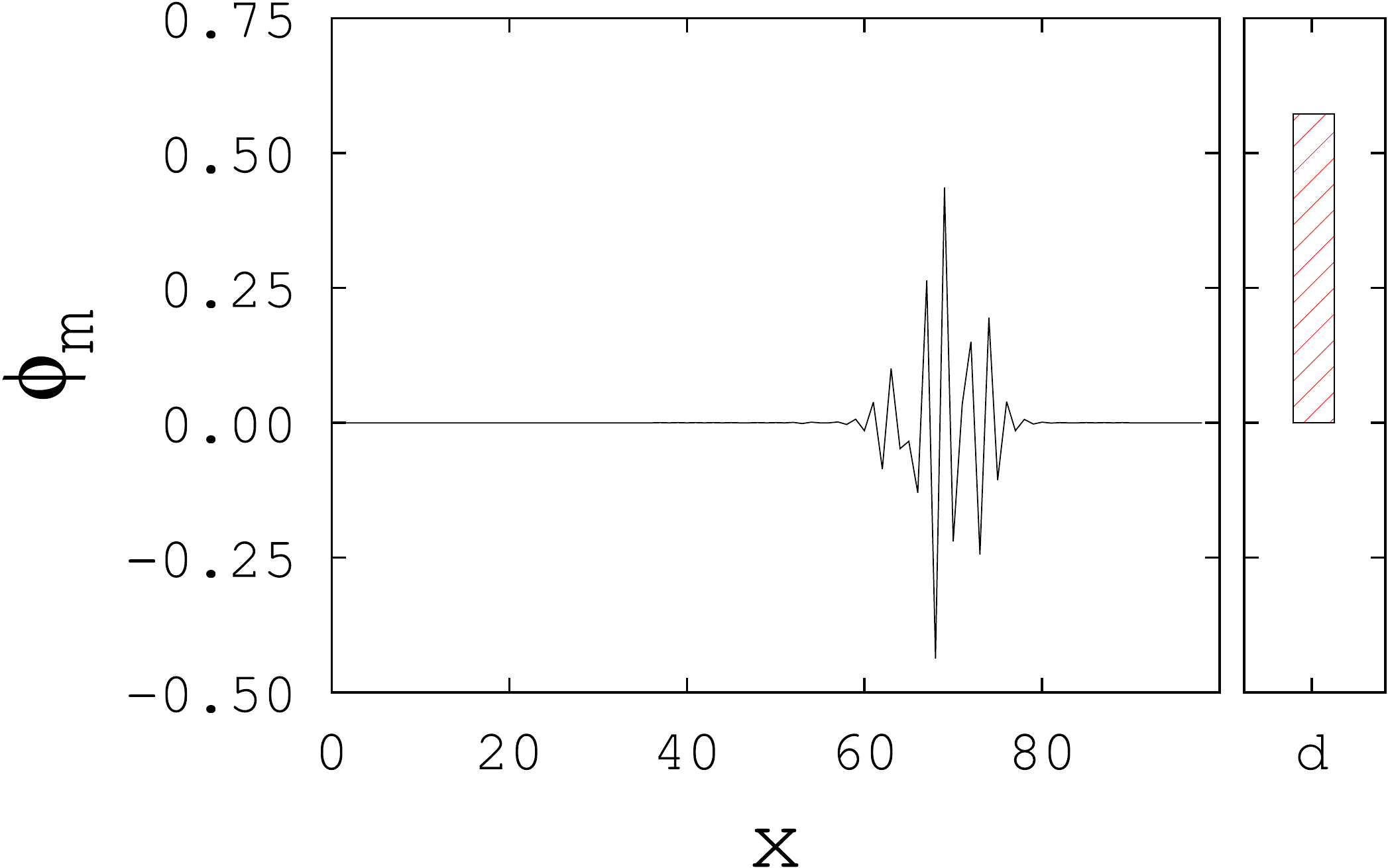}
                \caption{}
                \label{PAL89}
        \end{subfigure}%
        ~ %add desired spacing between images, e. g. ~, \quad, \qquad, \hfill etc.
          
        \begin{subfigure}[h]{0.5\textwidth}
                \includegraphics[width=2in, height=3.25in, angle=-90]{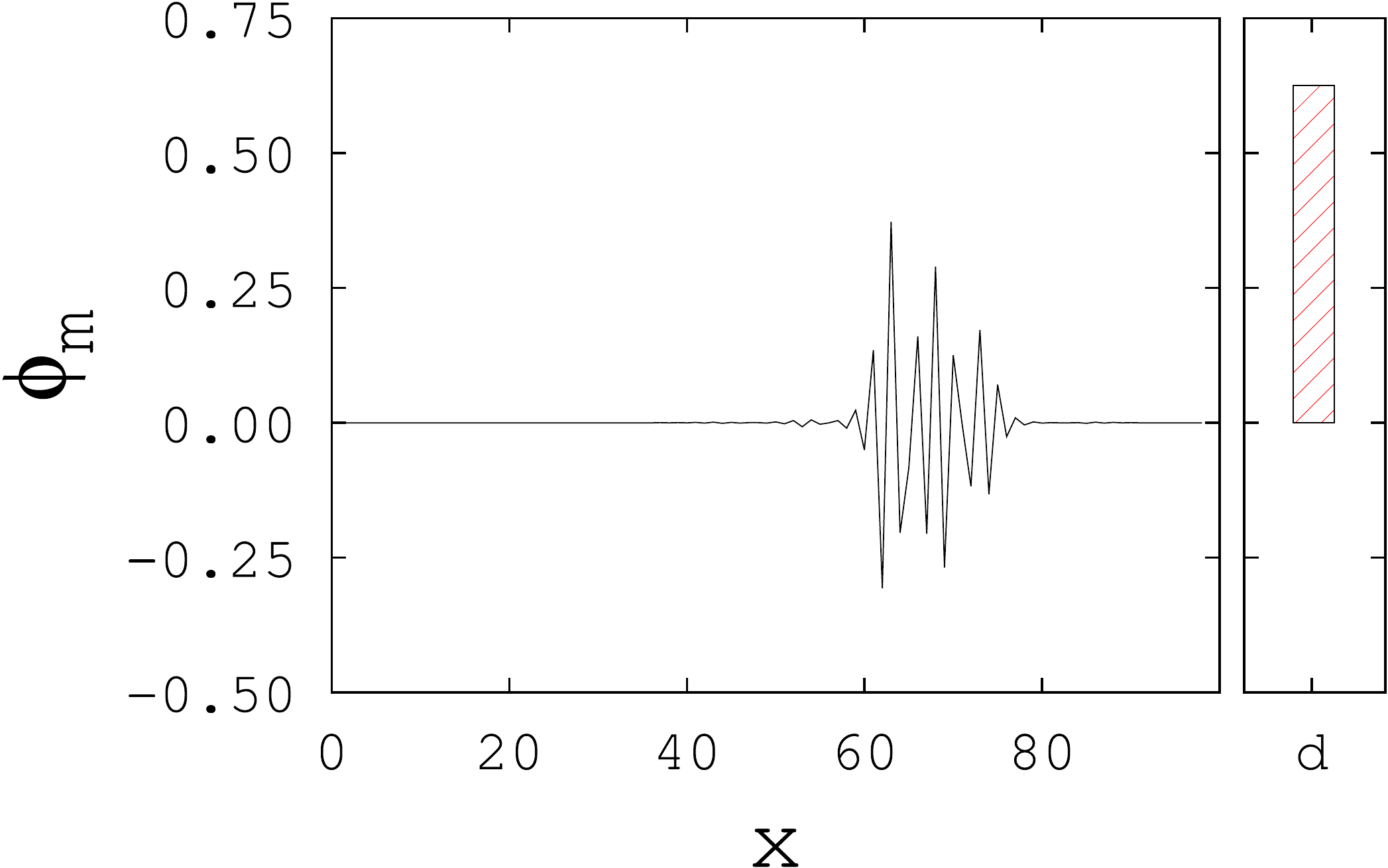}
                \caption{}
                \label{PAL91}
        \end{subfigure}
        ~ %add desired spacing between images, e. g. ~, \quad, \qquad, \hfill etc.
             \begin{subfigure}[h]{0.5\textwidth}
                \includegraphics[width=2in, height=3.25in, angle=-90]{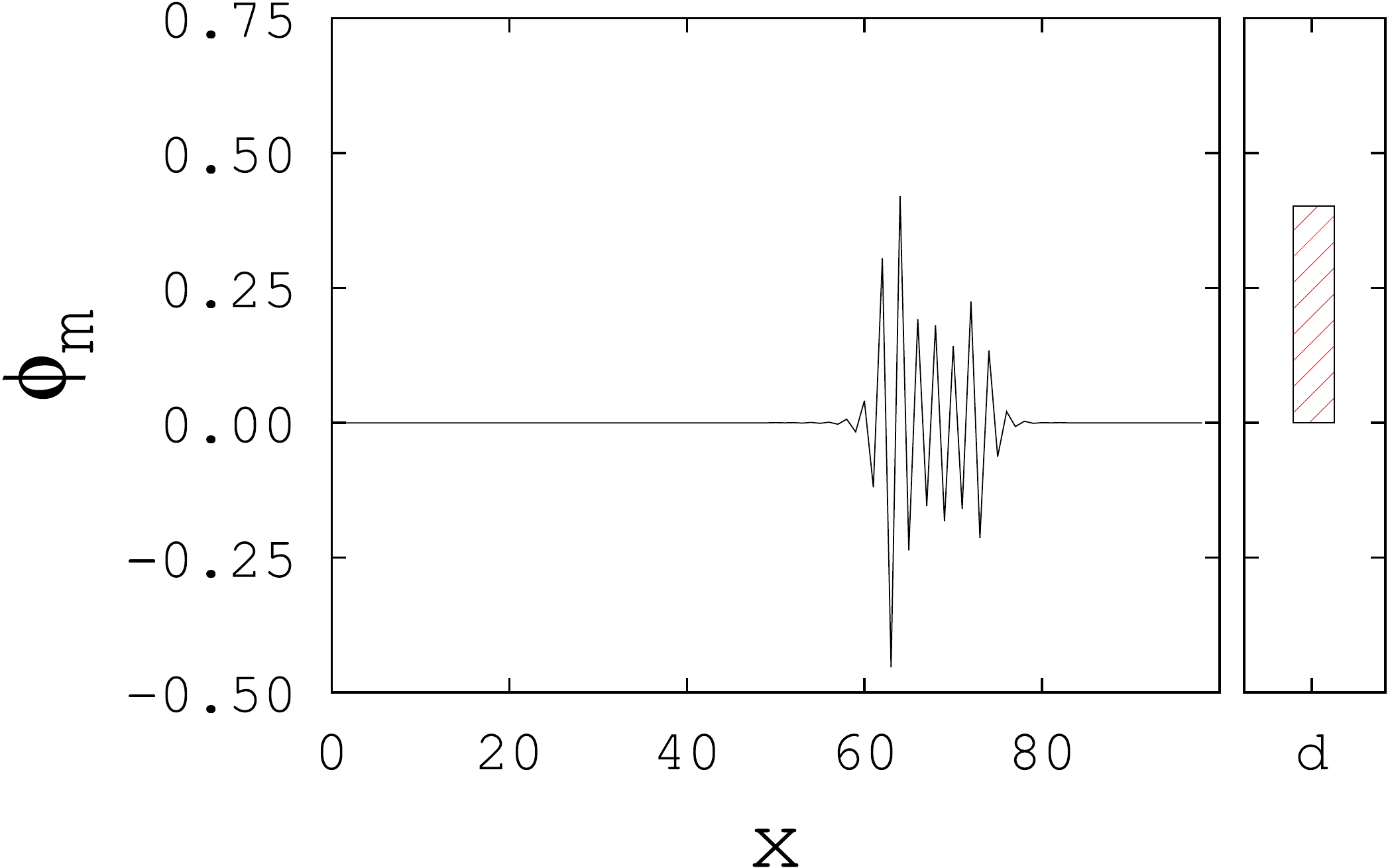}
                \caption{}
                \label{PAL97}
        \end{subfigure}
        ~ %add desired spacing between images, e. g. ~, \quad, \qquad, \hfill etc.
        \caption{Strongly hybridized state amplitudes within our numerical model for the $g=b/4$, $a=66$ and $\epsilon_\imp=-0.62$ case.  The side bar shows the amplitude of each state at the impurity site. (a) A hybridized AL state near maximum hybridization with the impurity state. The unperturbed form of this state is shown in Fig. \ref{AL0}. (b) The  impurity state hybridized with  the AL state in Fig. \ref{AL0} and (to a lesser extent) with another AL state similar to the state shown in (c).  (c) An AL state that is less hybridized with the impurity state than (a). }
\label{pert}
\end{figure}

Strongly hybridized states that include both lattice sites and the impurity site are shown in Fig. \ref{pert}.
In this figure, we have used the same specific realization of disorder as in Fig. \ref{IP}. 
We chose the attachment site $a=66$ because at this site there appears a sharply localized state. We chose $g=b/4=0.25$ because we found that on average it led to the hybridization of just a few AL states for different values of $\epsilon_\imp$, simplifying our analysis. For example, for  $\eps_\imp=-0.62$  only $2$ AL states are significantly hybridized with the impurity state. This is the value of $\eps_\imp$ used in Fig. \ref{pert}.

    Hybridization of AL states is manifested by the existence of a nonzero amplitude at the impurity site, while hybridization of the impurity state  is manifested by the existence of nonzero amplitudes on the lattice sites.  Hybridized states will enable control over spectral properties of the lattice and thus control of transport of the electron between the lattice and the impurity.

%plot3
\captionsetup[subfigure]{justification=raggedright, singlelinecheck=on}

%%plot 2
%%%%%%%%%%%
%The derivative of the perturbed eigenvalue with respect to the unperturbed one is related to the normalization constant as
%%
%\begin{eqnarray}
%\frac{\partial z_\kk}{\partial E_\kk} = N_\kk^2
%\end{eqnarray}
%%

%%
%%%%%%%%%%%%%%%%%
\section{Avoided crossings and hybridization}
\label{sec:avoid}
%%%%%%%
We will show that the coupling between the impurity and the attachment site leads to the hybridization of the unperturbed impurity state with a set of unperturbed lattice states; the latter  can be chosen by varying $\epsilon_\imp$. The resulting hybridized states are eigenstates of the full Hamiltonian. 

\begin{figure}[h]
\begin{center}
\includegraphics[width=3in, height=3.25in, angle=-90]{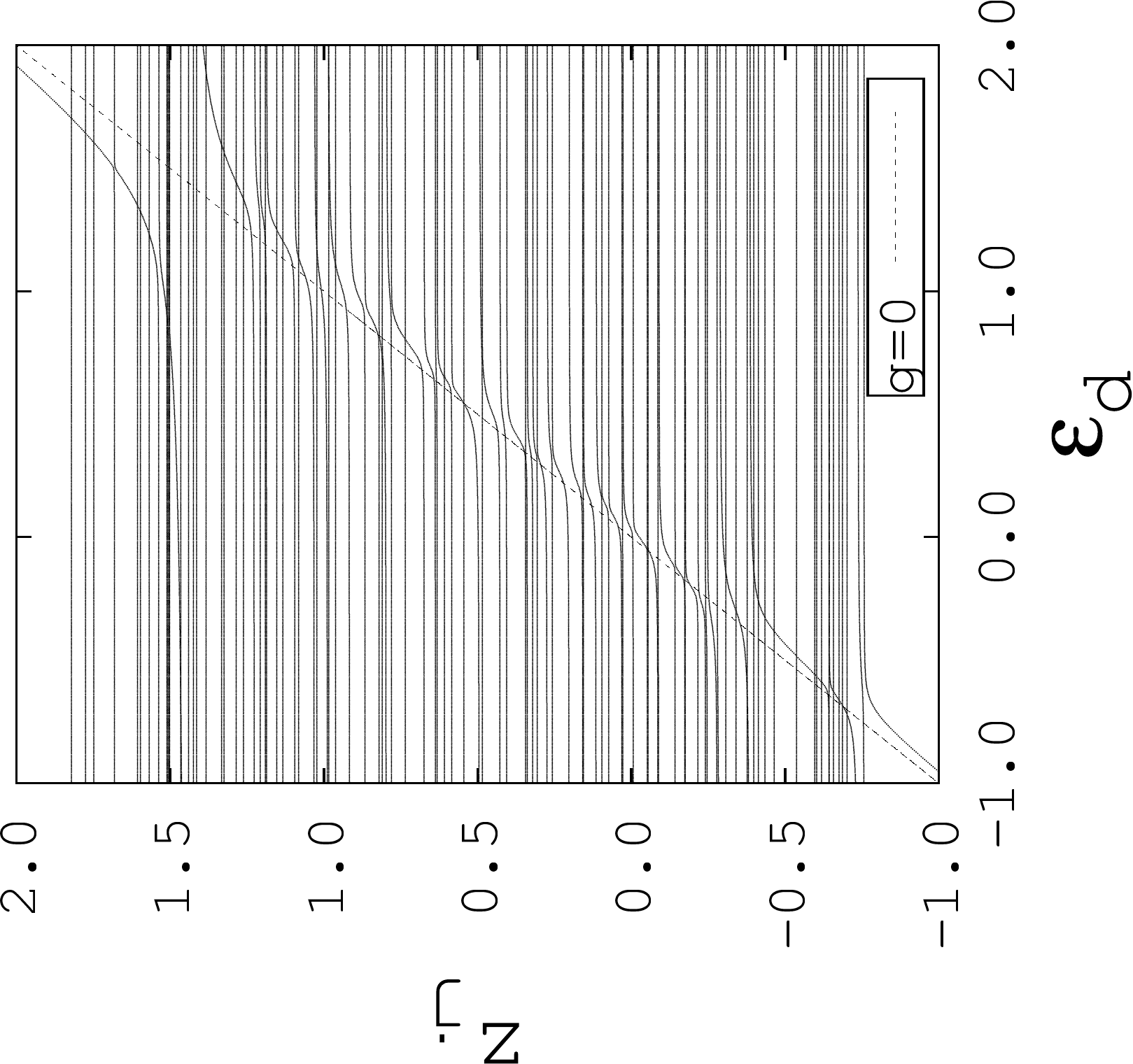} 
\end{center}
\caption{The eigenvalues $z_j$ of our numerical model are plotted against impurity energy $\epsilon_\imp$ for $a=66$ and $g=b/4$ in Fig. \ref{TJL}. The hybridized impurity state eigenvalue trends close to its $g=0$ value (dotted line) while the hybridized AL state eigenvalues are represented by the approximately horizontal plots.}
%We find that $\epsilon_\imp$ has strong control over the perturbed impurity  eigenvalue (Eq. (\ref{zev})) while limited control over perturbed AL eigenvalues (Eq. (\ref{zkev})).}
\label{spec}
\end{figure}

To investigate how  $\epsilon_\imp$  may select lattice states to become hybridized,   we start by numerically computing eigenvalues for the complete Hamiltonian  at different values of $\epsilon_\imp$. Figure \ref{spec} demonstrates the results using the previously mentioned parameter values and the specific set of random energies used in Fig. \ref{IP}. 
%%%%%%%%%%%
%%%%%%%%%%

%%%%%%%%
\begin{figure}[h]
\begin{center}
\includegraphics[width=2.5in, height=3.25in, angle=-90]{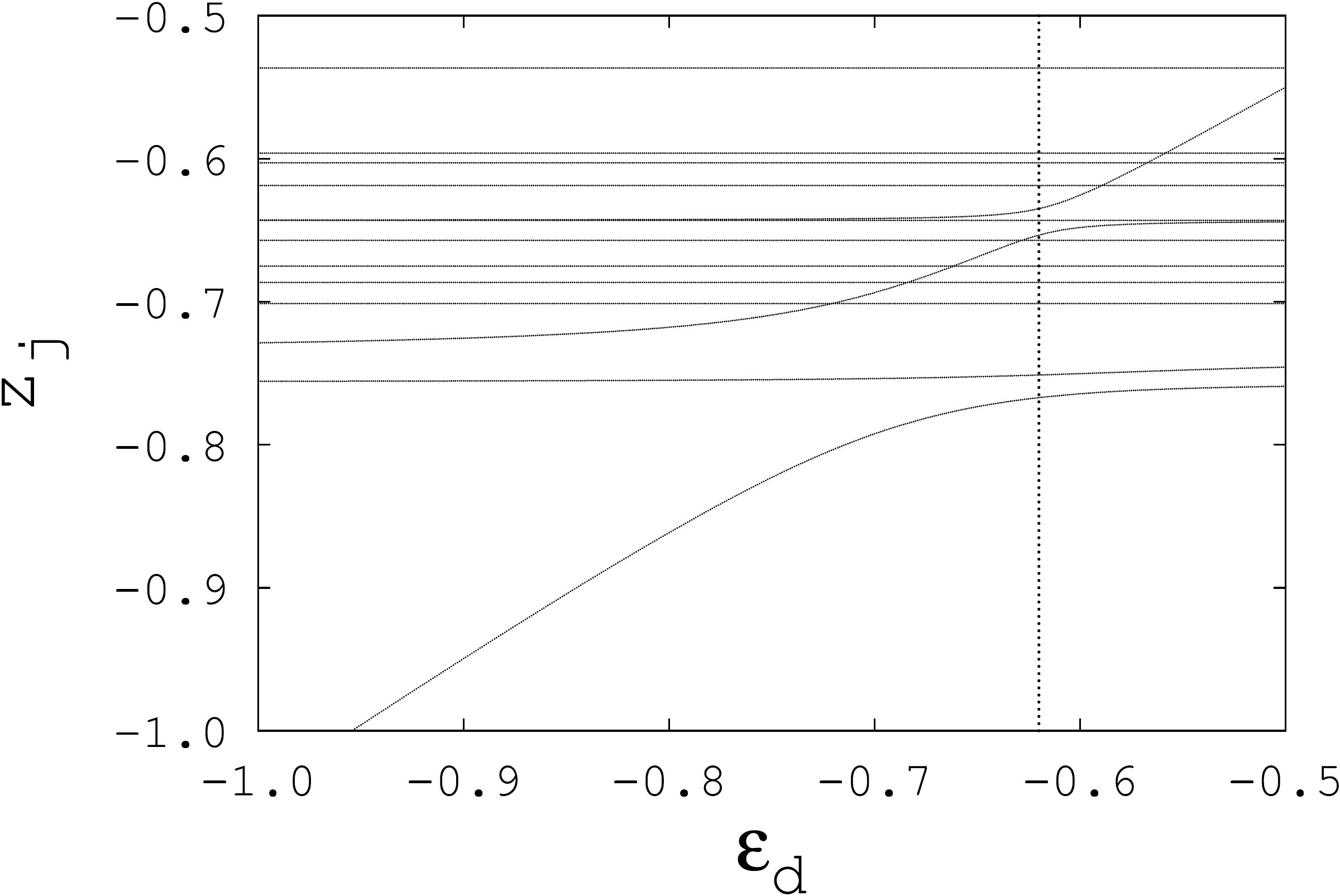} 
\end{center}
\caption{A portion of the lattice spectra is shown in detail for $a=66$, $g=b/4$, and varying $\epsilon_\imp$. As $\epsilon_\imp$ is increased from $-1.0$ to $-0.5$ two AL states become hybridized with the impurity state, which is indicated by the two visible avoided crossings. The perturbation of these states at $\epsilon_\imp=-0.62$ (vertical dashed line) is shown in Fig. \ref{pert}.}
\label{spec88}
\end{figure}
%%%%%%%%%%%%

If the impurity were uncoupled to the lattice (i.e. if $g=0$) then Figure \ref{spec} would show a set of $N$ horizontal lines, corresponding to the eigenvalues of the uncoupled lattice Hamiltonian; these eigenvalues are independent of $\epsilon_\imp$. Figure \ref{spec} would also show a diagonal line (indicated by the dashed line), corresponding to the eigenvalue $\epsilon_\imp$ of the uncoupled impurity state. 

When $g\ne 0$, some of the uncoupled eigenvalues are noticeably perturbed; the perturbation  is manifested in Figure \ref{spec} as avoided crossings, consisting of curved lines near the diagonal. Figure  \ref{spec88}  shows some of the avoided crossings in more detail.  As we will argue next, this perturbation of eigenvalues implies strong hybridization of the corresponding eigenstates.

 Eq. (\ref{phid3}) shows that for each perturbed eigenstate $|\phi_j\ket$, the probability 
\begin{eqnarray}
p_j \equiv \left| \bra \imp|\phi_j\ket\right|^2\end{eqnarray}
to find the electron at the impurity state $|\imp\ket$ is given by the slope of the curve of $z_j$ vs. $\epsilon_\imp$.  Away from the (visible) avoided crossings in Figures \ref{spec} or  \ref{spec88}, the horizontal lines have a slope near $0$ and thus $p_j\approx 0$; they correspond to lattice states with negligible perturbation. Meanwhile, the diagonal lines have a slope near $1$ and thus $p_{j'}\approx 1$; they correspond to the impurity state with  a small  perturbation. 
 
 As we approach an avoided crossing, following the curve associated with one of the lattice eigenvalues (nearly horizontal line) from the left,  the slope $p_{j}$ of the curve increases, while the slope $p_{j'}$ (nearly diagonal line) decreases. This means that  the probability to find the electron at the impurity shifts from the perturbed impurity state to the perturbed lattice state. At the middle point of the avoided crossing the slopes of the curves are approximately equal, thus $p_{j} \approx p_{j'}$.  Moving to the right, away from the avoided crossing, the eigenstates switch curves and $p_{j}$ decreases while $p_{j'}$ increases. Therefore, the middle of the avoided crossing is a point of maximum sharing of probability at the impurity site; it is a point of maximum hybridization. We identify the {\it degree of hybridization} of the two states with the product $p_{j} p_{j'}$.  In the next section, we will show that in the simplest case, maximum hybridization indeed occurs when the slopes are equal at the middle point of the avoided crossings. 

Note that $\sum_j p_j=1$. This means that if only one AL state $j$ is significantly hybdridized with the impurity state, then at maximum hybridization we have  $p_j=p_{j'}  = 1/2$ and the maximum degree of hybridization is $1/4$. If more than one AL state is hybridized, then we will have that $p_j<1/2$ and $p_{j'}<1/2$; the maximum degree of hybridization between any two states is then less than $1/4$. 

An example of this situation, with two significantly hybridized AL states,  is seen in Figure \ref{spec88}. It  occurs at the point indicated by the vertical dashed line. The hybridized states involved in the upper avoided crossing are maximally hybridized; these are the states shown in Figures \ref{PAL89} and \ref{PAL91}. Notice that the impurity site amplitude values are nearly equal, thus validating our previous statements. The lower avoided crossing in figure \ref{spec88} overlaps with the upper avoided crossing and involves the hybridized state  in Figure \ref{PAL97}. Notice that this state has a smaller amplitude at the impurity site, corresponding to a smaller slope of the bottom curve in Fig. \ref{spec88}.

While the attachment site and the width of disorder can alter the AL states available for perturbation, the impurity energy determines which available AL state(s)  become hybridized as shown in figure \ref{spec}. Thus we find that $\epsilon_\imp$ is an effective way to control perturbation within the lattice. Because there are degrees of hybridization, as indicated by avoided crossings in Fig. \ref{spec}, we find that $\epsilon_\imp$ can be used to tune maximum hybridization between AL state(s) and the impurity state. 

%%%%%%%%%%%
\section{Maximum Hybridization}
\label{sec:maxh}
%%%%%%%%%%
For some values of the impurity energy $\eps_\imp$ only one of the AL states (say the $m^{\rm th}$ state) is significantly hybridized with the impurity state due to the coupling $g$.  Assuming this is the case, in this section we show that

\smallskip
i) Maximum hybridization between the impurity state and the  AL state occurs when $\eps_\imp = E_\kk$, to zeroth order in $g$. 

\smallskip
ii)  At maximum hybridization the difference (gap) between the impurity and AL eigenvalues   across an avoided crossing  is a minimum; this minimum value is given  by $|2 g V_\kk|$.

iii) At the point of maximum hybridization the slopes of the curves of the eigenvalues vs. $\eps_\imp$ are equal.

\smallskip
iv) Partial hybridization between the impurity state and the AL state occurs when $|\eps_\imp-E_\kk| \lesssim 2g |V_\kk|$.

\smallskip
In more general cases, several AL states can be hybridized simultaneously. For these cases the results presented here are only rough approximations, applicable to the AL state that is most hybridized.  

To demonstrate (i-iv), we start by writing Eq. (\ref{zev}) as
\begin{eqnarray}\label{zev2}
z_j =  \eps_\imp  +  \sum_{m'\ne \kk} \frac{g^2|V_{m'}|^2}{z_j-E_{m'}}
+  \frac{g^2|V_\kk|^2}{z_j-E_\kk}.
\end{eqnarray}
Defining
\begin{eqnarray}\label{zev23}
{\tilde\eps}_{\imp,m}(z_j) \equiv \eps_\imp + \sum_{m'\ne m} \frac{g^2|V_{m'}|^2}{z_j-E_{m'}},
\end{eqnarray}
Eq. (\ref{zev2}) is re-written as
\begin{eqnarray}\label{zev25}
z_j = \frac{1}{2}\left({\tilde\eps}_{\imp,m}(z_j)+ E_\kk \pm \sqrt{({\tilde\eps}_{\imp,m}(z_j) - E_\kk)^2  + 4 g^2 |V_\kk|^2} \right) . \nonumber\\ 
\end{eqnarray}
Let us assume that ${\tilde\eps}_{\imp,\kk}(z_j)$ is approximately independent of $z_j$, and is approximately equal to $\eps_\imp$. This occurs if all the unperturbed AL states other than the $m^{\rm th}$ state have an eigenvalue $E_{m'}$ sufficiently far from $z_j$, such that the summation in Eq. (\ref{zev23}) is negligible.  In this case we have (with the labeling $j=\pm$)
\begin{eqnarray}\label{zev26}
z_\pm = \frac{1}{2}\left(\eps_\imp+ E_\kk \pm \sqrt{(\eps_\imp - E_\kk)^2  + 4 g^2 |V_\kk|^2} \right),
\end{eqnarray}
which gives a simplified description of the avoided crossings seen in Figs. \ref{spec} and \ref{spec88}. In the limit  $g\to 0$ we can see that $z_+$ gives the impurity energy $\eps_\imp$ and $z_-$ gives the AL energy $E_\kk$. For $g\ne 0$ the two solutions correspond to the two perturbed states resulting from the hybridization of the unperturbed impurity state and the $m^{\rm th}$ unperturbed AL state. 

Maximum hybridization occurs when the product of the probabilities of the two hybridized states at the impurity site is maximum. From Eq. (\ref{phid3}) this implies that 
\begin{eqnarray}\label{zev27}
\frac{\partial}{\partial\eps_\imp}\left(\frac{\partial z_+}{\partial\eps_\imp} \frac{\partial z_-}{\partial\eps_\imp}\right) = 0,
\end{eqnarray}
or
\begin{eqnarray}\label{zev277}
\frac{\partial}{\partial\eps_\imp} \left(\frac{g^2|V_\kk|^2}{(\eps_\imp-E_\kk)^2 + 4g^2|V_\kk|^2}\right)= 0,
\end{eqnarray}
which gives $\eps_\imp=E_\kk$ and the minimum distance 
\begin{eqnarray}\label{zev28}
z_+-z_-=2g|V_\kk|
\end{eqnarray}
 between the two hybridized eigenvalues. When $\eps_\imp=E_\kk$ we also have that the slopes of $z_\pm$ vs. $\eps_\imp$ are equal:
\begin{eqnarray}\label{zev288}
\left.\frac{\partial z_+}{\partial\eps_\imp}\right|_{\eps_\imp=E_\kk}  = \left.\frac{\partial z_-}{\partial\eps_\imp}\right|_{\eps_\imp=E_\kk}= \frac{1}{2}.
\end{eqnarray}

 Partial hybridization occurs when the term inside parenthesis in Eqs. (\ref{zev27}) or (\ref{zev277}) is non-negligible. This happens roughly when 
\begin{eqnarray}\label{zev29}
|\eps_\imp-E_\kk| \lesssim 2g |V_\kk| .
\end{eqnarray}
When $|\eps_\imp-E_\kk| = 2 | g V_\kk|$ the product of slopes in Eq. (\ref{zev27}) takes half its maximum value.

Finally, note that when $\eps_\imp=E_\kk$, the perturbed eigenvalues in Eq. (\ref{zev26}) are $z_\pm=E_\kk\pm g|V_\kk|$, which agrees with the previously stated condition (\ref{sigVk}) for significant 
interaction between the impurity and the lattice (i.e. for hybridization of eigenstates).
%%%%%%%%%%%%%
\section{Control of electron transport}
%%%%%%%%%%%%%%%

So far we have discussed how the impurity's energy can be tuned to alter the spectrum of the lattice and form hybridized states localized at  both lattice and impurity sites.  Here we demonstrate how this tuning can be used to direct electron transport within the disordered lattice. 

We begin  by considering the time evolution of our system for fixed values of $\epsilon_\imp$. The initial condition is  a single electron placed at the impurity at $t=0$. We consider the survival probability that the electron remains at the impurity at time $t$; time is defined in units of $b^{-1}$ with $\hbar=1$. Beginning with the electron at the impurity will produce an evolving superposition of perturbed states as time progresses. 

\begin{figure}[h!]
\centering
        \begin{subfigure}[b]{0.5\textwidth}
        \centering
                \includegraphics[width=2.5in, height=3.25in, angle=-90]{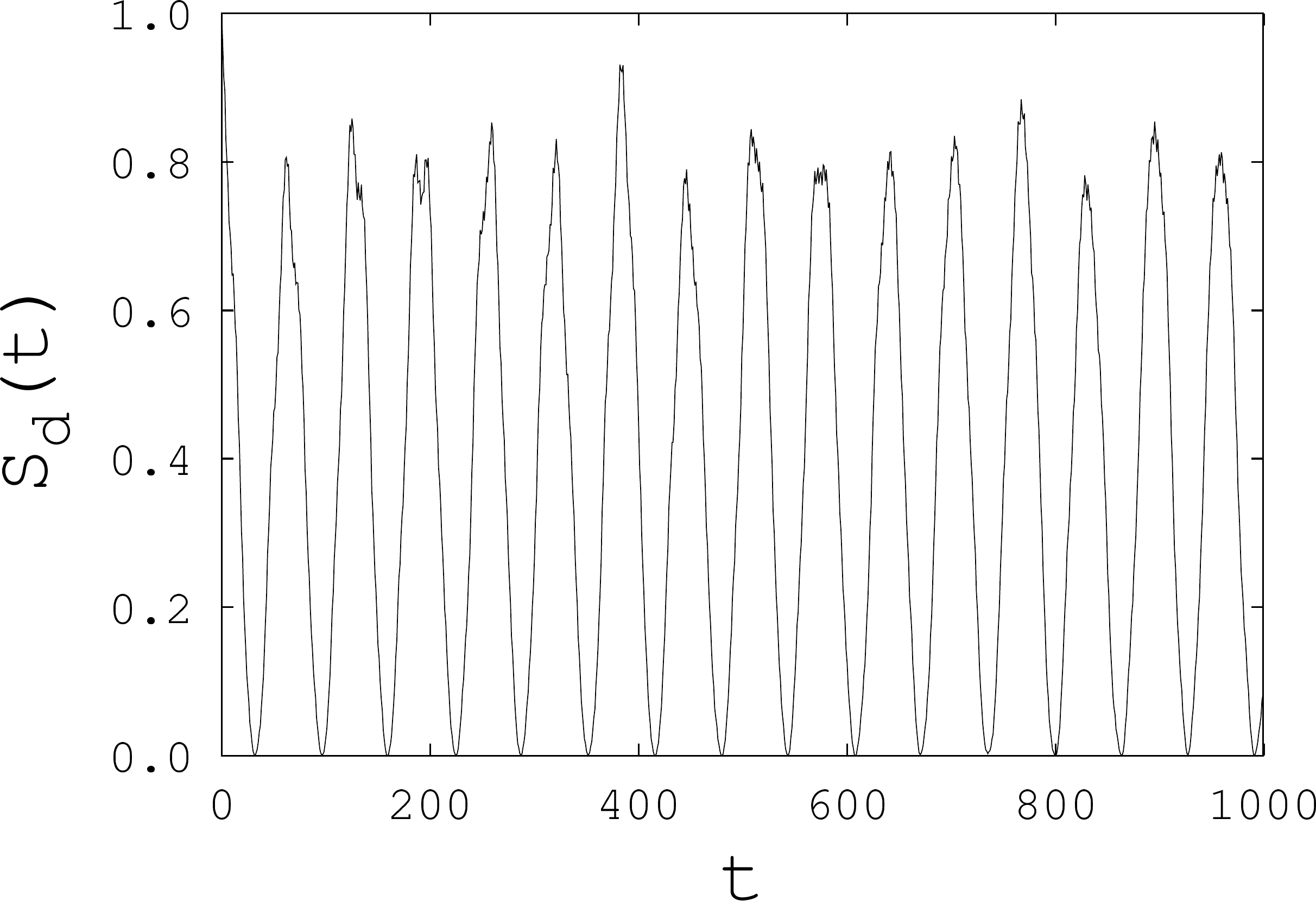}
                \caption{}
                \label{srabi}
        \end{subfigure}%
        ~ %add desired spacing between images, e. g. ~, \quad, \qquad, \hfill etc.
          
        \begin{subfigure}[b]{0.5\textwidth}
        \centering
                \includegraphics[width=2.5in, height=3.25in, angle=-90]{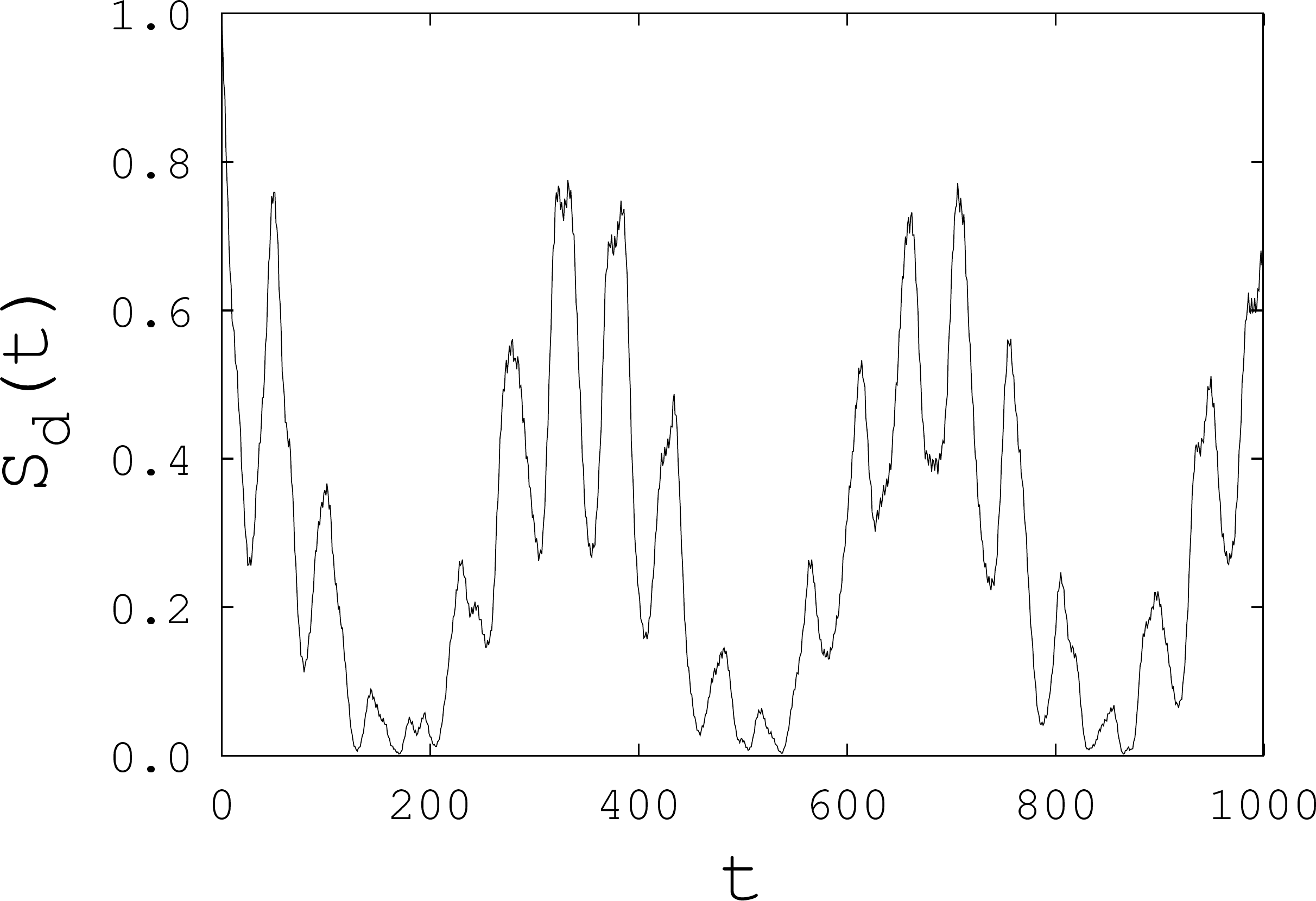}
                \caption{}
                \label{drabi}
        \end{subfigure}
        ~ %add desired spacing between images, e. g. ~, \quad, \qquad, \hfill etc.

%\begin{subfigure}[b]{0.5\textwidth}
%        \centering
%                \includegraphics[width=2.5in, height=3.25in, angle=-90]{far_rabi.pdf}
%                \caption{}
%                \label{far_rabi}
%        \end{subfigure}

        \caption{Survival probability that a single electron remains at site $\imp$ at time $t$ for $a=66$ and fixed $\eps_\imp$. (a) For $\epsilon_\imp=-0.69$ the  profile at site $\imp$ shows mainly one degree of periodicity indicating that only one hybridized AL state and the hybridized impurity state are involved in the superposition. (b) For $\epsilon_\imp=-0.62$ the profile at site $\imp$ shows mainly two degrees of periodicity indicating that two perturbed AL states and the perturbed impurity state are involved in the superposition. However, only one AL state has reached maximum hybridization with the impurity state. Minimum values of zero demonstrate that the electron has temporarily left the impurity site.}
\label{sevolve}
\end{figure}

The survival probability for the electron to remain at the impurity is expressed as
\begin{eqnarray}\label{surv1}
S_\imp(t) = \left| \bra \imp|e^{-iH t}  |\imp\ket\right|^2.
\end{eqnarray}
Using the complete set of eigenstates of the Hamiltonian $\{ |\phi_n\ket \}$ with eigenvalues $z_n$, we have
\begin{eqnarray}\label{surv2}
S_\imp(t) &=& \left|\sum_{n=1}^{N+1} \left| \bra \imp | \phi_n\ket \right|^2 e^{-iz_n t}\right|^2  %\nonumber\\
%&=&  |\sum_{n=1}^{N+1}  \frac{\partial z_n}{\partial\eps_\imp} e^{-iz_n t}|^2 
\end{eqnarray}
%%
%where in the second line Eq. (\ref{phid3}) was incorporated. 
Naturally,  the survival probability is dominated by eigenstates that have a large probability at the impurity site, which are the strongly hybridized states. 
Evolving our lattice in time will introduce a phase difference between these states, leading to oscillations, which will allow for dynamic electron transport as time progresses. In the simplest case, discussed in the previous section, where only one AL state is hybridized with the impurity state, the superposition of the two hybridized states  produces oscillations with period
\begin{eqnarray}\label{surv22}
T=\frac{2\pi}{|z_+-z_-|},
\end{eqnarray}
which at maximum hybridization (Eq. (\ref{zev28})) gives
\begin{eqnarray}\label{surv222}
T=\frac{\pi}{|g V_\kk|}.
\end{eqnarray}
As shown in Fig. \ref{srabi}, we numerically verified the existence of these oscillations  for the case of maximum hybridization, for which the oscillations of the survival probability resemble Rabi oscillations. The figure shows that the  minimum values of the survival probability are nearly zero. These minimum values are important because they demonstrate an instant in time in which the electron has completely left the impurity  and is instead located within the lattice. We can understand this as due to a destructive interference between hybridized eigenstates, such as the ones shown in Figure \ref{pert}, whose similar amplitudes at the impurity  add with opposite phase and cancel each other out. Thus by maximizing hybridization we are able to momentarily confine the electron in the lattice. The periodicity of the survival probability also allows electron transport to be predictable.
%%%%%
%%
%%
%%%%%%%%%%%%%%%%%%%%%%%%%%%%%%
\begin{figure}[h!]
\centering
        \begin{subfigure}[h]{0.5\textwidth}
        \centering
                \includegraphics[width=2.5in, height=3.25in, angle=-90]{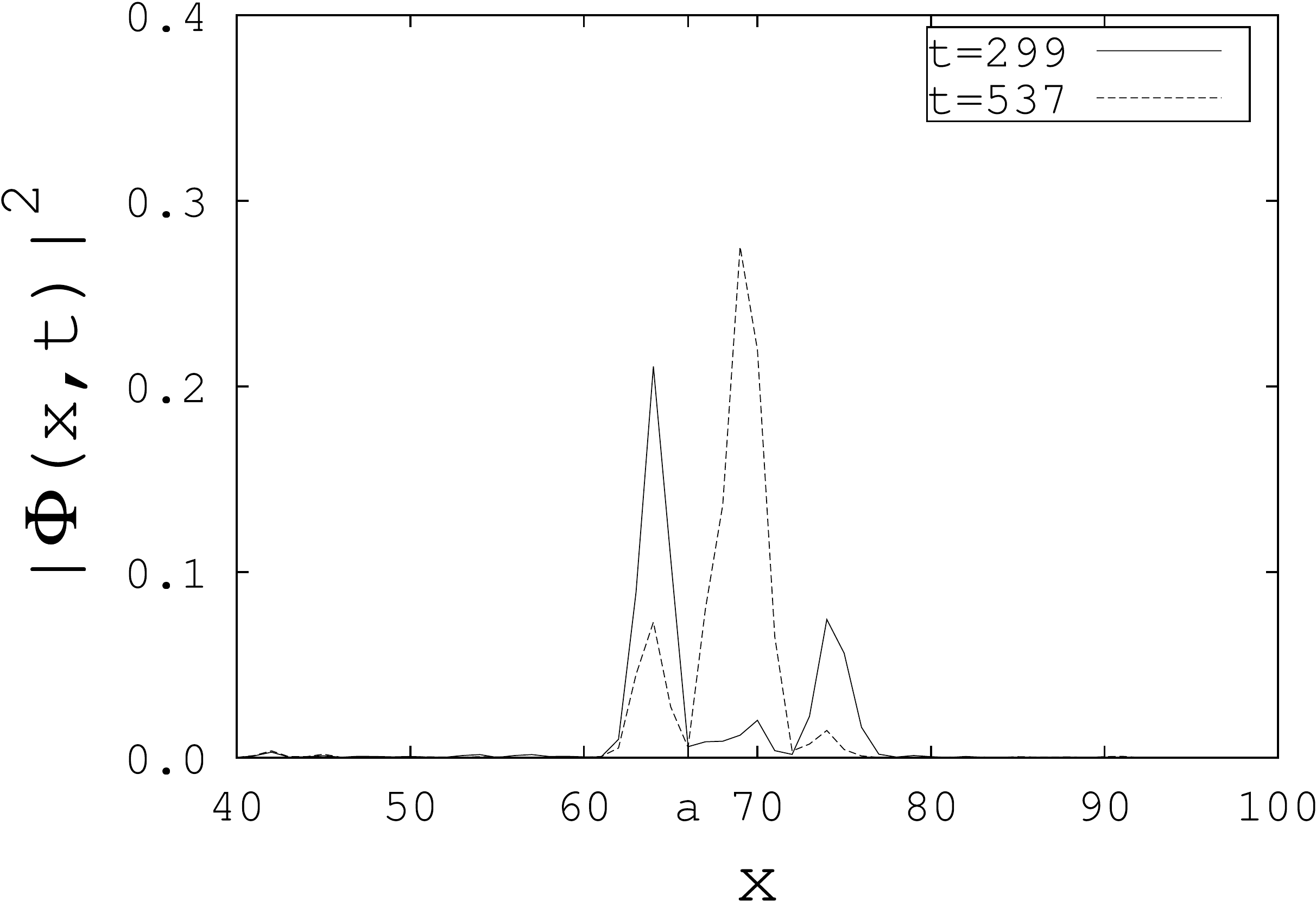}
                \caption{}
                \label{post}
        \end{subfigure}
        
        \begin{subfigure}[h]{0.5\textwidth}
        \centering
                \includegraphics[width=2.5in, height=3.25in, angle=-90]{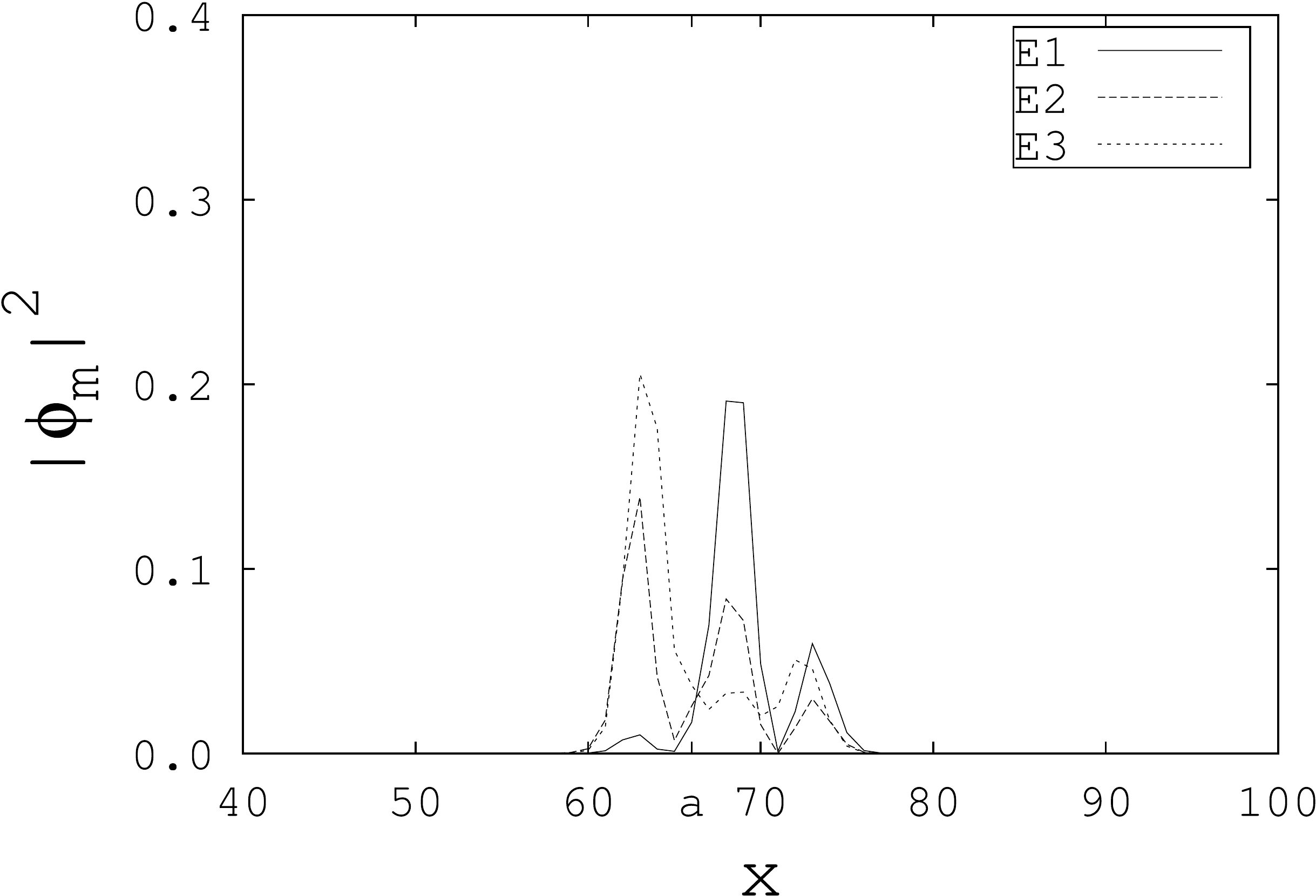}
                \caption{}
                \label{posk}
        \end{subfigure}%
        ~ %add desired spacing between images, e. g. ~, \quad, \qquad, \hfill etc.
        \caption{The above figures demonstrate the similarities between time evolved probability and perturbed lattice eigenstates. (a) Distribution of time-evolved probability across lattice sites for $\eps_\imp=-0.62$ and fixed $t$. The attachment site is indicated by ``a''. (b) Eigenstates of our numerical model for $a=66$, $g=b/4$, and $\epsilon_\imp=-0.62$. The curves $E_1$, $E_2$, and $E_3$ are the probability (amplitude squared) vs. position corresponding to the wave functions in figures \ref{PAL89}, \ref{PAL91}, and \ref{PAL97} respectively. }\label{pos}
\end{figure}
%%%%%%%%%%%%%%%%%%%%

The degree of periodicity in the Rabi oscillation is determined by how many perturbed AL states are involved in the perturbation. For instance, the profile of Fig. \ref{srabi}, corresponding to $\eps_\imp=-0.69$,  has essentially one degree of periodicity demonstrating that only one perturbed AL state (in addition to the perturbed impurity state) is significantly involved in the lattice perturbation. In this instance the state shown in Fig. \ref{PAL97} has reached maximum hybridization. This can be visualized in the spectrum at the center of the lower avoided crossing in Fig. \ref{spec88}, around $\eps_\imp=-0.69$.

Figure \ref{drabi}, corresponding to $\epsilon_\imp=-0.62$, has essentially two degrees of periodicity and thus two perturbed AL states involved in addition to the perturbed impurity state (a third AL state produces an additional long periodicity with very small amplitude, which is not visible in Figure \ref{drabi}). The energy $\epsilon_\imp=-0.62$ corresponds to the vertical dashed line in Figure \ref{spec88}. The states involved in the Rabi oscillation are those shown in Figure \ref{pert} and correspond to eigenvalue curves with non-negligible slope at the intersection with the vertical dashed line in Figure \ref{spec88}.

Although in  general our numerical model exhibits Rabi oscillations with higher  s of periodicity, we are primarily interested in low degrees of periodicity, which enables greater transport control.

To better understand where the electron  is located during the minimum and local minimum points of the survival probability in Figure \ref{drabi}, we consider the probability of the  electron to be at any lattice site $x$ at given time $t$,
\begin{eqnarray}\label{probx}
|\Phi(x,t)|^2 = \left| \bra x|e^{-iH t}  |\imp\ket\right|^2.
\end{eqnarray}
Figure \ref{post} compares how the time evolved probability is distributed amongst lattice sites at a local minimum ($t=299$ from figure \ref{drabi}) against that at an absolute minimum point ($t=537$). Figure \ref{posk} shows the spatial probability distribution of the three hybridized states that participate most strongly  in the time evolution. We label them as $E_1$, $E_2$ and $E_3$.  These states correspond to the wave functions of figures \ref{PAL89}, \ref{PAL91}, and \ref{PAL97} respectively. State $E_1$ is a maximally hybridized AL state; $E_2$ is the hybridized impurity state and $E_3$ is a partially hybridized AL state. At the absolute minimum point ($t=537$) the states $E1$ and $E2$ interfere destructively at the impurity site, but they interfere constructively in the lattice. Hence the probability distribution in Fig. \ref{post} resembles a superposition of $E1$ and $E2$ in Fig. \ref{posk}. 

At the local minimum  ($t=299$)   Fig. \ref{post} resembles $E3$ in Fig. \ref{posk} because then $E1$ and $E2$  approximately cancel in the lattice. The similarities between figures \ref{posk} and \ref{post} demonstrate what we would expect from our previous analysis; when the electron is not at site $\imp$ it is at locations that are roughly defined by the localization of the perturbed AL states $E_1$ and $E_3$ and the  perturbed impurity state $E_2$.

Thus we have demonstrated the impurity energy's ability to control electron transport within our finite disordered lattice. By tuning $\epsilon_\imp$ we can force the electron to oscillate between the impurity site and specific groups of localized sites as time progresses.

%%%%%%%%%%%%%%%%
\section{Tuning the range of electron oscillations}
\label{sec:Tuning}
%%%%%
\begin{figure}[h!]
\centering      
                \includegraphics[width=2.5in, height=3.25in, angle=-90]{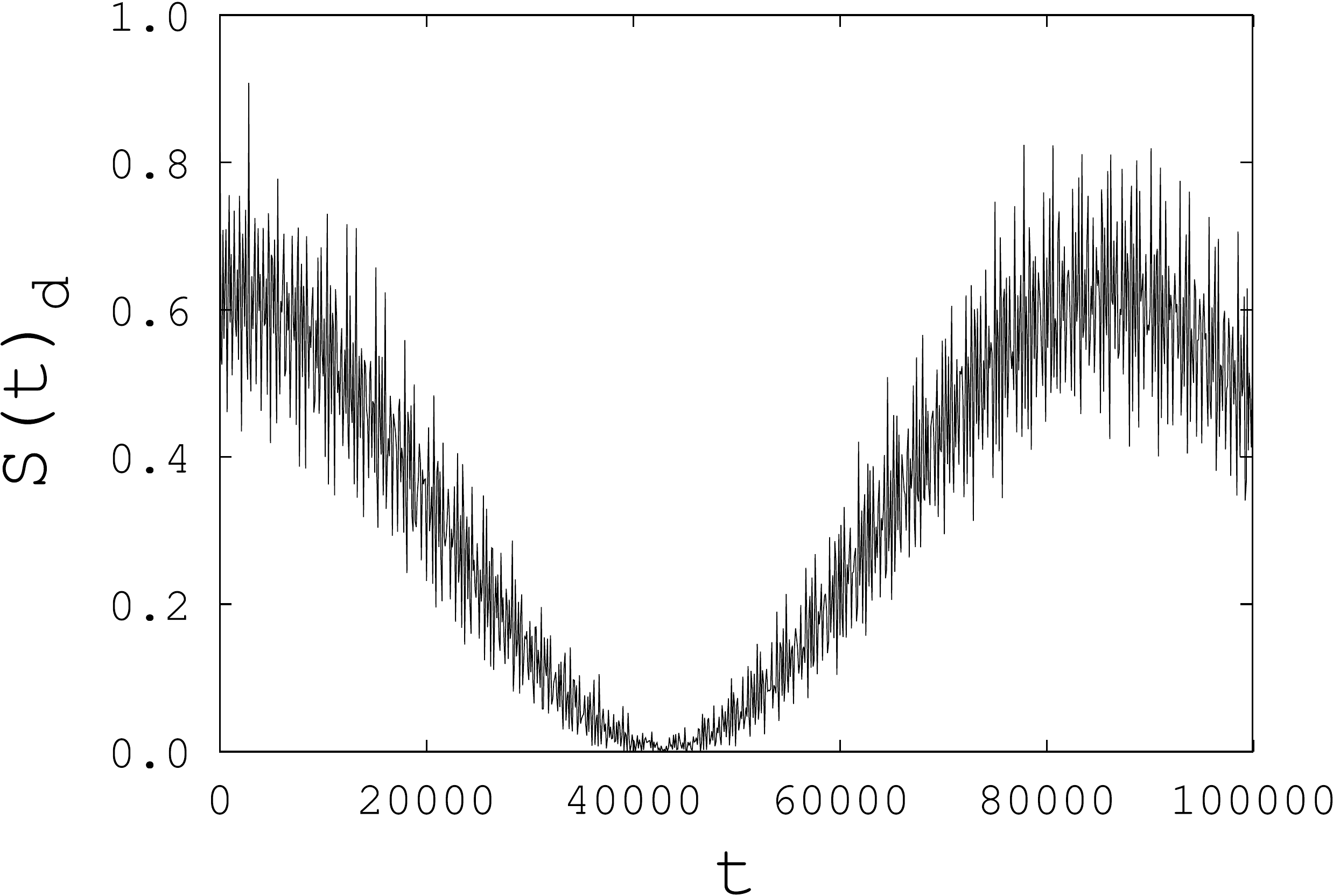}     
        \caption{Survival probability that a single electron remains at site $\imp$ at time $t$ for $a=66$,
        % $\epsilon_\imp=-0.4830379796483107$%
        $\epsilon_\imp=-0.48304$ and $g=0.25$. To attain significant hybridization $\epsilon_\imp$ should be within the range $E_\kk \pm2 g |V_\kk|$. In this case $V_\kk=1.6\times10^{-4}$, which gives the range $E_\kk \pm 8\times 10^{-5}$. The  profile at site $\imp$ shows mainly one degree of periodicity indicating that mainly one hybridized AL state and the hybridized impurity state are involved in the superposition. The period of the oscillation is about $80000$, in approximate agreement with Eq. (\ref{surv222}).}
\label{periodf}
\end{figure}

The electron oscillations described in the previous section involved AL states that are close to the attachment site. Therefore the spatial range of the oscillations was limited to the Anderson localization length of these states. 

However, in general, it is possible to select values of $\eps_\imp$ that  allow the impurity state to hybridize with an AL state far from the attachment site. To see this, consider that an AL state with a given inverse localization length $\gamma(E_\kk)$  has an amplitude given approximately by the exponential function
\begin{eqnarray}\label{ALdist0}
\psi_\kk(x) \approx \sqrt{\gamma(E_\kk)} e^{-\gamma(E_\kk)|x-x_0|}
\end{eqnarray}
where $x_0$ is the point of localization (maximum amplitude). Given that $\psi_\kk(a)=V_\kk$, we obtain 
\begin{eqnarray}\label{ALdist}
V_\kk\approx \sqrt{\gamma(E_\kk)} e^{-\gamma(E_\kk)D}
\end{eqnarray}
 where $D\equiv |a-x_0|$ is the distance from the point of the localization of the AL state to the attachment site. This means that $V_\kk$ decreases exponentially with $D$. However, it is possible to tune $\eps_\imp$ so that it is at the center of the avoided crossing formed by the AL and the impurity eigenvalues. In this case the corresponding states will be maximally hybridized and we will have a Rabi oscillation, where the electron travels back and forth between the impurity and the region around $x_0$. In order for this hybridization to occur, $\eps_\imp$ must be within the range $E_\kk\pm 2g |V_\kk|$ as shown in Eq. (\ref{zev29}). Therefore as $D$ increases, $\eps_\imp$ has to be tuned with a  precision that increases exponentially with $D$; the larger $D$ is, the smaller the gap at the avoided crossing. In Figures \ref{spec} and \ref{spec88} one can see such avoided crossings with very small gap; the resolution of these figures makes it seem that in many places the lines are simply crossing, but in fact these are all ``microscopic'' avoided crossings. These are each associated with long-range (large $D$) transport of the electron. 

From Eqs. (\ref{surv222}) and (\ref{ALdist}), the period of the oscillation at maximum hybridization is
\begin{eqnarray}\label{surv22D}
T=\frac{\pi}{g\sqrt{\gamma(E_\kk)}} e^{\gamma(E_\kk)D}
\end{eqnarray}
which increases exponentially with $D$ as well. Hence we find that the penetration distance $D$ into the lattice that the electron can achieve can be larger than the Anderson localization length. Meanwhile, the larger $D$ is the more precisely $\eps_\imp$ must be tuned, and the longer the period of oscillation becomes.

\begin{figure}[h!]
\centering
        \begin{subfigure}[h]{0.5\textwidth}
        \centering
                \includegraphics[width=2.5in, height=3.25in, angle=-90]{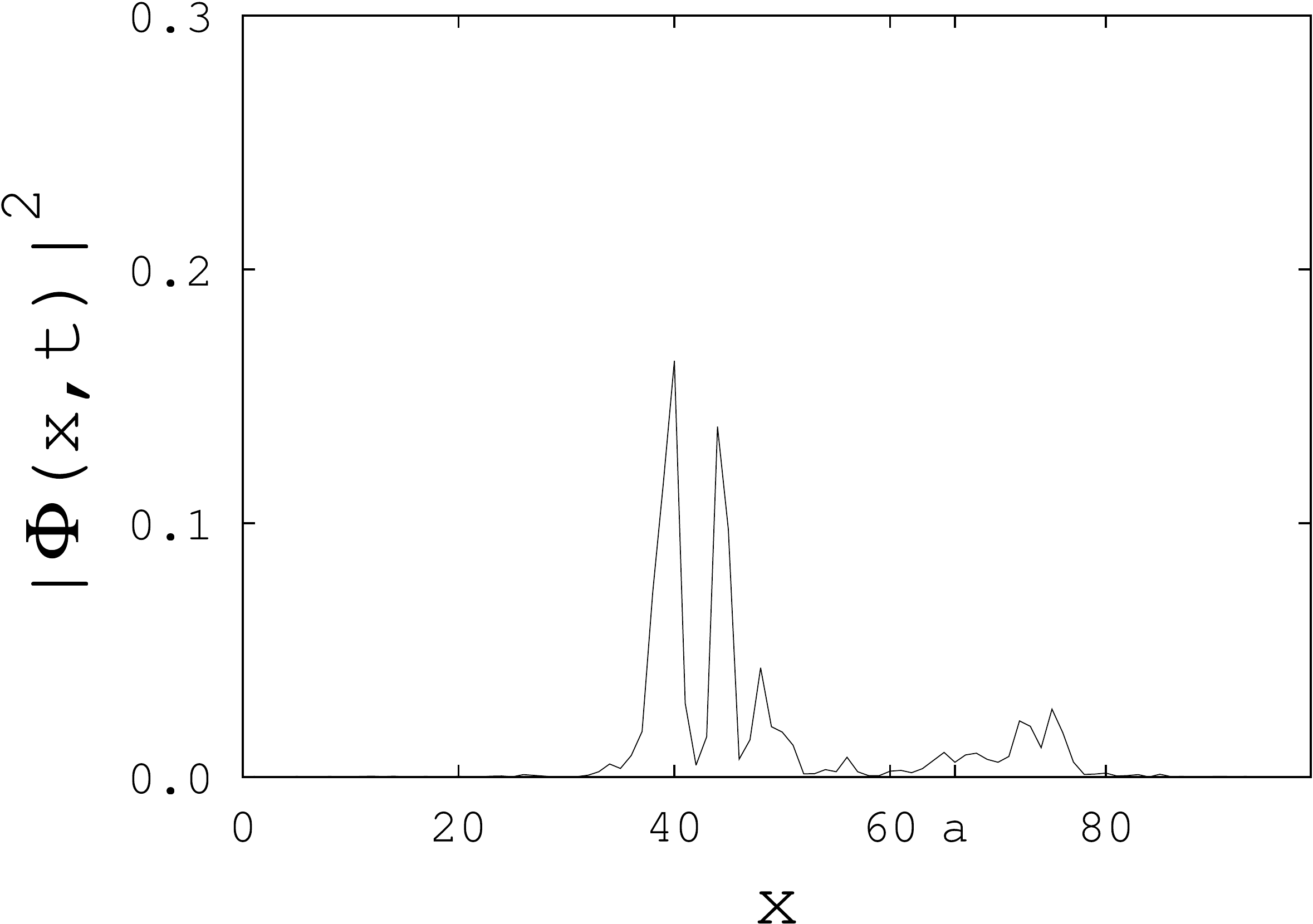}
                \caption{}
                \label{postf}
        \end{subfigure}
        
        \begin{subfigure}[h]{0.5\textwidth}
        \centering
                \includegraphics[width=2.5in, height=3.25in, angle=-90]{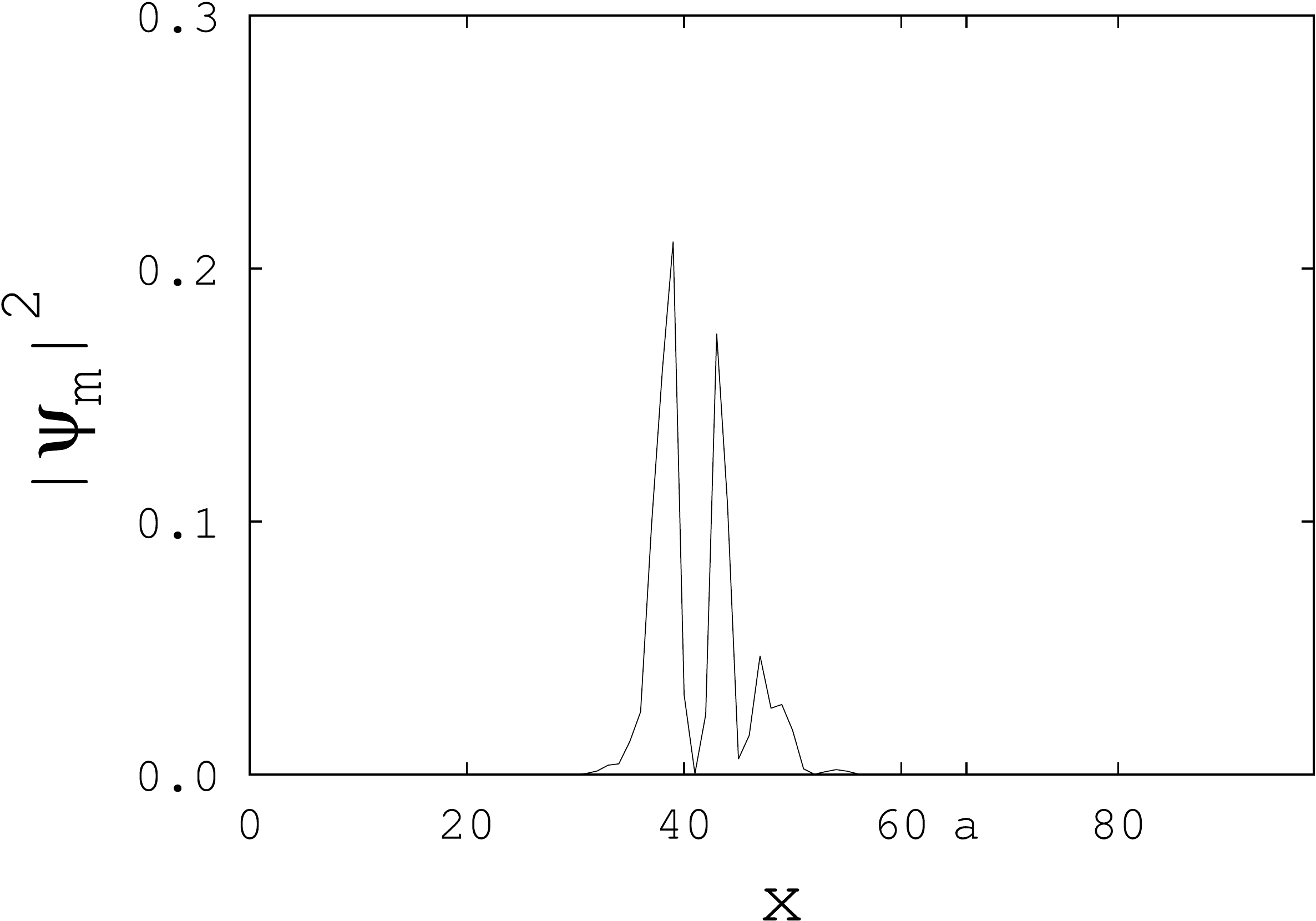}
                \caption{}
                \label{poskf}
        \end{subfigure}%
        ~ %add desired spacing between images, e. g. ~, \quad, \qquad, \hfill etc.
        \caption{ (a) Distribution of lattice time-evolved probability across lattice sites for $\eps_\imp=-0.48304$ and fixed $t=41000$, corresponding to a minimum of the survival amplitude in figure \ref{periodf}. The attachement site is indicated by ``a''. 
        (b) Uncoupled AL state (for $g=0$) that becomes strongly hybridized with the impurity state after the interaction is switched on for $g \neq 0$. The chosen $\eps_\imp=-.48304$ value is slightly off the center of the avoided crossing formed by the AL and impurity eigenvalues, but it is within the range of maximum hybridization mentioned in the text.  The probability to find the electron at the impurity is approximately $0.4$ for both the AL and impurity states after becomimg hybridized. This value is close to the theoretical maximum of $0.5$ discussed in Sec. \ref{sec:avoid}. 
        }\label{posf}
\end{figure}

If $D$ is not too large, achieving a medium-range transport is not very difficult; it  is enough that  $\eps_\imp$ lies within the narrow range of the avoided crossing. The electron transport is  illustrated in figures \ref{periodf} and \ref{posf}, which show, respectively, the survival probability as a function of time and the probability distribution of the wave function when the electron is farthest from the attachment site. For reference, Fig. \ref{posf}b shows the AL state that  becomes strongly hybridized with the impurity state.

%%%%%%%%%%%%%%%%%%%%%%%%%%%%%%
%%%%%%%%%%%%%%%%%%%%

%%%%%%%%%%%%%%%%%%%%%%%%%%%%%%%%%%%%%%%%%%%%%%%%%%%%%%%%%%%%%%%%%
\section{Ensemble Averaging}
\label{sec:Ensemble}

In this section we demonstrate that the transport properties we have considered thus far are robust with respect to ensemble averaging, which must be considered for modeling realistic lattice systems.
%To get a more clear picture of how the hybridized states can be used to control transport in a realistic disordered lattice, one should consider ensemble averaging.
This will allow us to find,  for a given impurity energy, the average number of  lattice states that are strongly hybridized with the impurity, as well as the average localization of those states. Since we want to focus on transport that is not affected by the boundaries of the lattice, we will also discuss the average number of hybridized states that are free from boundary effects.
%Most importantly, the ensemble averaging will show that the transport properties discussed in sections \ref{sec:TJ}-\ref{sec:Tuning} are valid not only for the specific realization of disorder we used thus far, but also for the average disordered lattice.

The number of strongly hybridized states can be qualitatively defined as the number of states with significant impurity-site amplitude. Quantitatively this can be determined as follows: for a specific realization of disorder, we first define the array of amplitudes at a site $x$ as 
\begin{eqnarray}\label{test}
{\cal A}_x\equiv \{\bra x|\phi_j\ket\}, \quad j=1\cdots N+1 .
\end{eqnarray}
The inverse participation number of this array is given by
\begin{eqnarray}\label{P_x}
P_x^{-1} \equiv \sum_{j=1}^{N+1} \left|\bra x|\phi_j\ket\right|^4 .
\end{eqnarray}
We then define the number of states that exhibit significant hybridization with the impurity state as the integer nearest to $P_d$, denoted as $n_d\equiv\left\lfloor{P_d}\right\rceil$. 
Thus we identify the $n_d$ states that exhibit significant impurity hybridization as the $n_d$ states with the highest amplitudes (absolute values) in the array ${\cal A}_d$. This set of amplitudes is written as 
\begin{eqnarray}\label{ad}
{\hat{\cal A}}_d \equiv \{n_d\, {\rm highest}\, \left|\bra d|\phi_j\ket\right|\}
\end{eqnarray}
%%

%It is also important to consider the number of hybridized states that are not influenced by the  boundaries of the lattice. We  define this influence quantitatively in terms of  the amplitude of the hybridized states at the end lattice sites.

%Having identified the hybridized states in Eq. (\ref{ad}), we next identify the subset of these states that  overlap with the end sites at $x=1$ and $x=N$.
However, in order to exclude boundary effects, we should identify the subset of
${\hat{\cal A}}_d$ that exhibit strong overlap with the end sites $x=1$ and $x=N$ and remove them from the larger set.
To do so we must form arrays of end-site amplitudes and calculate the participation number of these arrays, $P_{1}$ and $P_{N}$. After finding subsets ${\hat{\cal A}}_1$ and ${\hat{\cal A}}_N$ with the $n_{1} \equiv\left\lfloor{P_1}\right\rceil$ and $n_{N}\equiv\left\lfloor{P_N}\right\rceil$ highest amplitudes, respectively, we can then determine the number of eigenstates that significantly overlap both with the impurity site and with at least one end site.  These are the states belonging to the subset
\begin{equation}\label{aend}
{\cal A}_{d;{\rm ends}} = \\ {\hat{\cal A}}_d \cap{\hat{\cal A}}_1+  {\hat{\cal A}}_d\cap {\hat{\cal A}}_N
	.
\end{equation}
 We will denote the number of these states as $n_{d;{\rm ends}}$. Excluding these states from ${\hat{\cal A}}_d$ gives the hybridized states that are free from boundary effects.

 To find the ensemble average  of $n_d$, $n_{d;{\rm ends}}$, we did the following: keeping  the hopping energies and impurity position constant, we chose a standard error of the mean ($SEM$) threshold so that the size of the 95-percent confidence interval ($CI$) is equated to a desired fraction $f$ of the data-set mean $\mu$;  that is $CI=f \mu$, where the confidence interval is $(\mu-2\frac{SEM}{\sqrt{n_L}},\mu+2\frac{SEM}{\sqrt{n_L}})$, $n_L$ is the number of disordered  lattices in the ensemble and $CI=4\frac{SEM}{\sqrt{n_L}}$. We stopped averaging when the standard error of the mean fell below the threshold.

\begin{figure}[h!]
\centering      
                \includegraphics[width=3.25in, height=3in, angle=0]{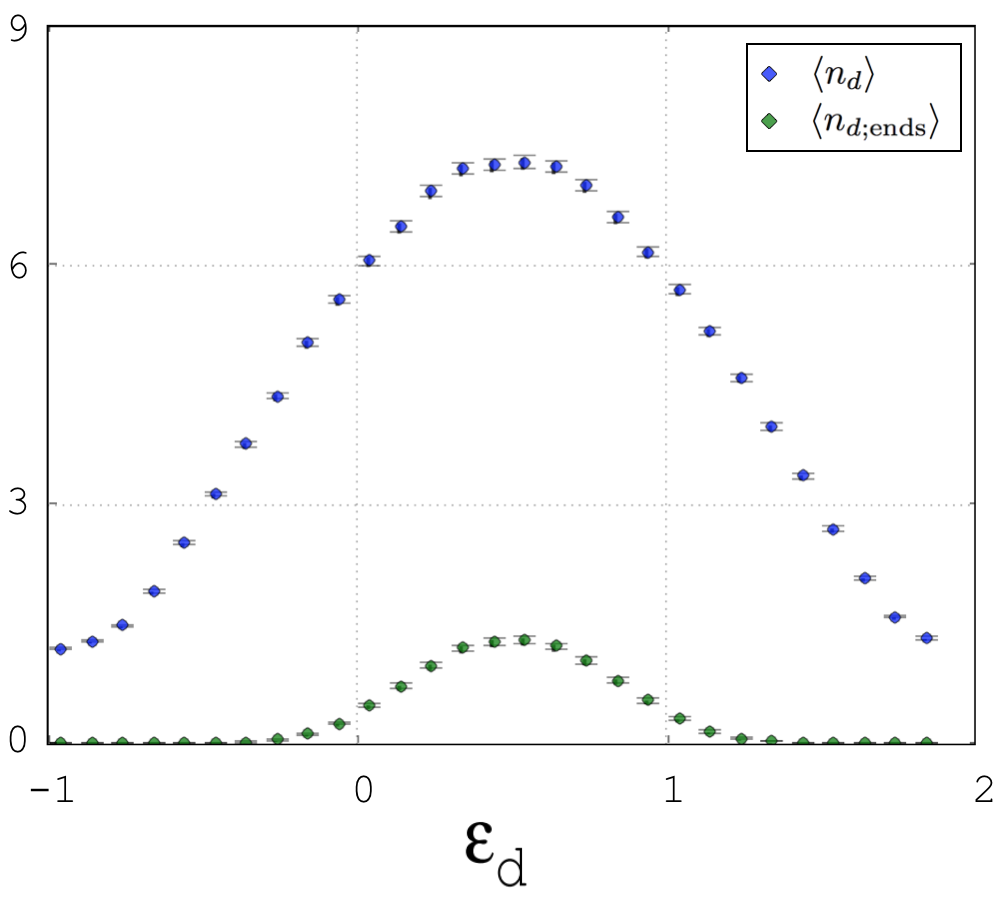}     
        \caption{Utilizing the averaging procedure described below Eq. (\ref{aend}), we plot the average number of hybridized states $\bra n_d\ket$ (upper blue curve, Eq. (\ref{ad})) and the number of hybridized states that overlap with the lattice edges $\bra n_{d;{\rm ends}}\ket$ (lower green curve, Eq. (\ref{aend})) for a range of impurity energies. The numerical calculation used $N=100$, $b=1$, $W=b$, $a=N/2$, $g=b/4$, confidence interval: $CI=0.01\mu$.}
\label{numHsysEsys}
\end{figure}

Figure \ref{numHsysEsys} presents the results of the above averaging procedure, which demonstrate that a side impurity can direct transport within any disordered lattice with little to no boundary influence. The upper curve in Figure \ref{numHsysEsys} gives the ensemble-averaged number of hybridized states $\bra n_d\ket$ vs. the impurity energy $\epsilon_d$, while the lower curve gives the average number of hybridized states  $\bra n_{d;{\rm ends}}\ket$ that have significant overlap with the ends of the lattice.  The regions $\epsilon_d\in[-1.0,0.0]$ and $\epsilon_d\in[1.0,2.0]$ are regions where we find hybridized states with minimal boundary influence. The average number of hybridized states within these regions varies  between approximately $1$ and $5$. Staying within these energy regions ensures that only states exhibiting strong localization interact with the impurity, which makes electron diffusion outside the strongly hybridized region unlikely. 

Note that there are two transition points
around $\epsilon_d \lesssim 0$ and $\epsilon_d \gtrsim 1$
where the lower curve in Fig.  \ref{numHsysEsys} begins to increase from zero or decrease to zero. These points correspond to a transition between hybridized states that are influenced and those that are not influenced by the lattice's finite size. Interestingly, at these points the slope of the upper curve changes slightly. 

Meanwhile, when the impurity has an energy outside the lattice energy spectrum $\epsilon_d < -1$ or $\epsilon_d > 2$ the total number of hybridized states approaches unity, indicating that there is only one state (the impurity state) with significant amplitude at the impurity.  This means the impurity  is isolated from the lattice. 
Naturally, the overlap of the impurity state with the endpoints also vanishes in this region of strong localization. 
%Meanwhile,   outside the  energy spectrum the number of hybridized states with end influence approaches zero. This is due to the high localization of the impurity state, which on average has negligible overlap with the end sites of the lattice, in this case. 

The average $\bra n_d\ket$ in Fig. \ref{numHsysEsys} roughly agrees with the results we obtained for the specific realization of disorder in sections~\ref{sec:TJ}-\ref{sec:Tuning}.  For example, for $\eps_d=-0.62$, we have $\bra n_d\ket\approx 3$ in Fig. \ref{numHsysEsys}, which means that in addition to the impurity state there are, on average, two hybridized lattice states.  This is consistent with the two avoided crossings seen in Fig. \ref{spec88} and  with the double oscillation in Fig. \ref{sevolve}b. 

%  Interestingly, when the lower curve in Fig.  \ref{numHsysEsys} begins to increase from zero or decrease to zero the slope of the upper curve changes accordingly. This corresponds to a transition between hybridized states that are influenced and those that are not influenced by the lattice's finite size.

%%%
%%%

In addition to the average number of strongly hybridized states, we also obtained their ensemble-averaged  inverse participation numbers, which are a measure of inverse localization length. We did this numerically by obtaining every hybridized state for a specific realization of disorder, as described in Eq.  (\ref{ad}), and averaging their individual inverse participation numbers. We then obtained ensemble averages of these average participation numbers,  following an averaging procedure similar to the one discussed earlier in this section. Figure \ref{IP} demonstrates the validity of our numerical calculations by showing an agreement between the inverse participation numbers of the eigenstates of a single disordered lattice, the theoretical expression for inverse localization length in a infinite disordered lattice (Eq. (\ref{gammaE})), and the ensemble-averaged  inverse participation number of hybridized states.

The numerical results in this section demonstrate that our method for impurity-directed transport in a disordered lattice is robust against ensemble averaging. Further, we can reliably predict the average number of hybridized states that are free from boundary effects. These states lead to a fairly regular oscillatory transport of the electron between the impurity and specific regions of the lattice. 

%%%%%%%%%%%%%%%%%%%%%%%%%%%%%%%%%%%%%%%%%%%%%%%%%%%%%%%%%%%%%%%%%
\section{Discussion}
\label{sec:disc}

We have considered a simple model for transport control within a finite disordered T-junction lattice. The presence of the side-coupled impurity significantly impacts the spectral properties of the lattice, forming  hybridized AL/impurity states. We have treated the  energy of the side impurity  as a tunable parameter that selects which states become hybridized,  thereby controlling the motion of the electron. In particular, the electron can be forced to periodically oscillate between the impurity site and specific groups of localized sites as time progresses.

%Although there are similarities between Pendry's investigation of finite disordered lattices \citep{Pendry87} and ours, there are clear distinctions as well. While Pendry focuses on transport through a finite disordered lattice, which relies on a ``necklace'' of several hybridized AL states that, by chance,  create a sub-band due to near-degeneracy, we focus on the transport from the impurity to a small region  of our finite lattice, by tuning the energy of the impurity to  select just a few AL states to become hybridized.

Previously Pendry \citep{Pendry87} has shown that transport can be achieved through a finite disordered chain by relying on hybridized AL states that form a ``necklace'' through such a lattice.  While this necklace of states forms a sub-band due to near-degeneracy that occurs by chance, in our model we demonstrate that a side-attached impurity can be used as a control device to intentionally select AL sites according to specific hybridization characteristics that enable desired transport properties.

Although we have studied a simplified model, our results may serve as a starting point for the design  of devices based on finite disordered lattices that attempt to  route electrons. Nano-designed transistors with disordered materials have already been proposed \cite{Guo}. Using disorder to trap electrons within a given area could help combat issues with device miniaturization. Moreover, lattices based on complex stacked T-junction structures may offer interesting possibilities for controlling or storing electrons. 

The finite size of disordered lattices we have considered may allow for nonzero temperature operation in realistic devices, because the energy-level spacings can be larger than the thermal excitation energies; this would allow the system to behave similarly to the zero-temperature case. Assuming that the width of the energy band is of the order of the width without disorder, the average level-spacing is  $2b/N$. A rough estimation of the maximum operational temperature would then be given by $k_BT=2b/N$. As a reference, for a disordered lattice with $N=100$, having $T=300 K$ would require $b=1.3$ eV. An issue that deserves further investigation is the following: The level spacings between hybridized states at avoided crossings can be much smaller than $2b/N$, so  thermal excitation could induce transitions between these states, similar to variable-range hopping. This would modify the simple Rabi oscillations discussed in the present paper for $T=0$. In Ref. \cite{Chomette} it was found experimentally that carrier transport in  a disordered superlattice becomes thermally activated around $T=77 K$.

Our results are applicable to disordered optical or microwave lattices, such as the microwave lattice of Ref. \cite{Bliokh} mentioned in the Introduction. For this type of lattice, temperature effects are much less important than for electron lattices, so our results regarding the Rabi oscillations and far-transport could be more readily tested on this type of system. 

A possible extension of our work is to make the impurity energy time-dependent and study the electron motion in this case. As shown in \cite{Levi} disordered optical systems systems with time-evolving disorder can produce ``hyper-transport'' of light, which is faster than ballistic transport. It would be interesting to see if this can be achieved in the case of electron transport. Related to this,  Ref. \cite{Sahel} investigates transport driven through a disordered lattice by applying time-dependent control fields at the edges of the lattice.

\acknowledgements
We thank Naomichi Hatano, Satoshi Tanaka and Kenichi Noba for insightful
discussions. We thank the Institute of Industrial Science at the University of Tokyo, the Katrina Roch Seitz Science Education Fund, the Holcomb Awards Committee and the LAS Dean's office at Butler University for support of this work. S. G. acknowledges support from a Young Researcher's Grant from Osaka Prefecture University. 

{}

\begin{thebibliography}{}


\bibitem{A1} G. D. Mahan, {\it Many Particle Physics}, 2nd Ed., Plenum Press, New York (1990).

\bibitem[Orellana (2003)]{Orellana03} P. A. Orellana, F. Dominguez-Adame, I. Gomez and M. L. Ladron de Guevara, Phys. Rev. B \textbf{67}, 085321 (2003).

\bibitem{B1} A. Dhar, D. Sen, and D. Roy, Phys. Rev. Lett. {\bf 101}, 066805 (2008).

\bibitem{A2} D. Roy, Phys. Rev. B {\bf 80}, 245304 (2009).
\bibitem{B2} A. Nishino, T. Imamura, and N. Hatano, Phys. Rev. Lett. {\bf 102}, 146803 (2009).
\bibitem{A3} S. Garmon, H. Nakamura, N. Hatano, and T. Petrosky, Phys. Rev. B {\bf 80}, 115318 (2009).
\bibitem{A4} S. Tanaka, S. Garmon and T. Petrosky, Phys. Rev. B 73, 115340 (2006).
\bibitem{A5} V. Eisler and S. Garmon, Phys. Rev. B {\bf 82}, 174202 (2010).

\bibitem[Sasada (2011)]{Sasada11} Keita Sasada, Naomichi Hatano, and Gonzalo Ordonez, J. Phys. Soc. Jpn. \textbf{80}, 104707 (2011).


\bibitem[Anderson (1958)]{Anderson58} P. W. Anderson, Phys. Rev. \textbf{109}, 1492 (1958).


\bibitem{Kramer}  B. Krammer and A. MacKinnon, Rep. Prog. Phys \textbf{56},  1469 (1993).
\bibitem{Lagendijk} A. Lagendijk, B. van Tiggelen and D. S. Wiersma, Physics Today p. 24, (Aug.  2009).
\bibitem{Segev} M. Segev, Y. Silberberg and D. Christodoulides, Nature Photonics {\bf 7}, 197 (2013).
\bibitem{Hsieh} P. Hsieh et al,  Nature Physics {\bf 11}, 268 (2015).
\bibitem{Chomette} A. Chomette, B. Deveaud, A. Regreny, and G. Bastard
Phys. Rev. Lett. {\bf 57}, 1464 (1986).
\bibitem[Landauer (1957)]{Landauer57} R. Landauer, IBM J. Res. Dev. \textbf{1} 223 (1957).

\bibitem[Landauer (1970)]{Landauer70} R. Landauer, Phil. Mag. \textbf{21} 863 (1970).

\bibitem[Erdos (1982)]{Erdos82} P. Erdos and R. C. Herndon, Adv. Phys. \textbf{31} 65 (1982).

\bibitem[Lambert (1984)]{Lambert84} C. J. Lambert, Phys. Rev. B \textbf{29}, 1091 (1984).
\bibitem[Kappus (1981)]{Kappus81} M. Kappus and E. Wegner, Z. Phys. B \textbf{45}, 15 (1981).
\bibitem[Pendry (1987)]{Pendry87} J. B. Pendry, J. Phys. C: Solid State Phys. \textbf{20}, 733-742 (1987).

\bibitem{Bertolotti} J. Bertolotti, S. Gottardo, D. S. Wiersma, M. Ghulinyan and L. Pavesi, Phys. Rev. Lett., {\bf 94} 113903 (2005).
\bibitem{Chen} L. Chen, W. Li and X. Jiang, New J. of Phys. {\bf 13}, 053046 (2011).
\bibitem{Kunz} H. Kunz and B. Shapiro, Phys. Rev. B {\bf 77} 054203 (2008).
\bibitem{Feinberg} J. Feinberg, Int J. Th. Phys. {\bf 50}, 1116 (2011).
\bibitem{PALee} P. A. Lee, Phys. Rev. Lett {\bf 53}, 2042 (1984).
\bibitem{Bosisio2} R. Bosisio et al.,   New J. Phys. {\bf 16} 095005 (2014).
\bibitem{A6} R. Notzel and K. H. Ploog, Adv. Mater. 5, {\bf 22} (1993).
\bibitem{A7} M. S. Gudiksen, L. J. Lauhon, J. Wang, D. C. Smith, and C. M. Lieber, Nature {\bf 415}, 617 (2002).

\bibitem{A8} K. Kobayashi, H. Aikawa, A. Sano, S. Katsumoto, and Y. Iye, Phys. Rev. B  {\bf 70}, 035319 (2004).
\bibitem{A9} T. Otsuka, E. Abe, S. Katsumoto, Y. Iye, G. L. Khym, and K. Kang, J. Phys. Soc. Jap. {\bf 76}, 084706 (2007).
\bibitem{Bliokh} K. Y. Bliokh, Y. P. Bliokh, V. Freilikher, A.Z. Genack and P. Sebbah, Phys. Rev. Lett {\bf 101} 133901 (2008).
\bibitem{Bosisio1} R. Bosisio, G. Fleury,
and J.-L. Pichard, New J. Phys. {\bf 16} 035004 (2014).


\bibitem[Thompson (2014)]{Thompson14} C. Thompson, Y. N. Joglekar, and G. Vemuri, arXiv:1404.3222v1 [physics.optics] (2014).

\bibitem[Wegner (1980)]{Wegner80} F. J. Wegner, Z. Phys. B \textbf{36}, 209 (1980).
\bibitem{Guo} X. Guo and S.R.P. Silva, Science {\bf 320} 618 (2008).
\bibitem{Levi} L. Levi et al, Nature Physics {\bf 8} 912 (2012).
\bibitem{Sahel} S. Ashhab, Phys. Rev. A {\bf 92}, 062305 (2015).


%^\bibitem[Ketzmerick (1998)]{Ketzmerick98} R. Ketzmerick, K. Kruse, and T. Geisel Phys. Rev. \textbf{80}, 137 (1998).

%\bibitem[Poddubny (2012)]{Poddubny12} Alexander N. Poddubny, et al, Nature Communications \textbf{3}, 914 (2012).

%\bibitem[Derrida (1984)]{Derrida84} B. Derrida and E. Gardner, J. Physique \textbf{45}, 1283 (1984).


%\bibitem[Garcia (2010)]{Garcia10} P. D. Garcia, et al, ArXiv \textbf{2}, 818 (2010).












\end{thebibliography}
\end{document}